\documentclass[%
 reprint,
superscriptaddress,
notitlepage,
nofootinbib,
 amsmath,amssymb,
 aps,
prd, 
floatfix,
]{revtex4-2}

\usepackage{graphicx}
\usepackage{dcolumn}
\usepackage{bm}

\usepackage{epsfig}
\usepackage{amsmath}
\input{epsf}
\usepackage{psfrag}
\usepackage{xcolor}

\newcommand{\bea}{\begin{eqnarray}}
\newcommand{\eea}{\end{eqnarray}}
\newcommand{\nn}{\nonumber}

\begin{document}

\title{Towards the Precision Nucleon Energy-Energy Correlator in Lepton-Ion Collisions}

\author{Haotian Cao}
\email{haotiancao@mail.bnu.edu.cn}
 \affiliation{Center of Advanced Quantum Studies, Department of Physics, Beijing Normal University, Beijing, 100875, China}

\author{Xiaohui Liu}
 \email{xiliu@bnu.edu.cn}
 \affiliation{Center of Advanced Quantum Studies, Department of Physics, Beijing Normal University, Beijing, 100875, China}
 \affiliation{Center for High Energy Physics, Peking University, Beijing 100871, China}

\author{Hua Xing Zhu}%
 \email{zhuhx@zju.edu.cn}
\affiliation{Zhejiang Institute of Modern Physics, Department of Physics, Zhejiang University, Hangzhou, 310027, China}%

\begin{abstract}
The nucleon energy-energy correlator (NEEC) was proposed in~\cite{Liu:2022wop} as a new way of studying nucleon intrinsic dynamics. In this work, we present a detailed derivation of the factorization theorem that enables the measurement of the unpolarized NEEC in lepton-ion collisions. As a first step towards a precise measurement of this quantity, we obtained the next-to-leading-logarithmic (NLL, $\sim{\cal O}(\alpha_s^n L^{n-1})$) resummation in a concise analytic form, and predicted the analytic $\theta$-angle distribution at ${\cal O}(\alpha^2_s)$. Extending our analytic resummation formula to higher logarithmic accuracy and the factorization theorem to hadron-hadron collisions is straightforward.
\end{abstract}

\date{\today}    
\maketitle

\section{Introduction}

Understanding the intricate internal structures of nucleons is at the central focus of nuclear physics for decades, and will continue to be the scientific frontier within the Standard Model at the next generation QCD facilities such as the upcoming electron-ion collider (EIC)~\cite{AbdulKhalek:2021gbh, Proceedings:2020eah,Anderle:2021wcy}. 
In recent years, our approaches to nucleon/nucleus tomography have been substantially enriched, thanks to the introduction of innovative ideas into the field, such as the jet-based studies of the transverse momentum dependent
(TMD) structure functions~\cite{Gutierrez-Reyes:2018qez,Liu:2018trl,Gutierrez-Reyes:2019msa,Gutierrez-Reyes:2019vbx,Arratia:2020nxw,Liu:2020dct,Arratia:2020ssx,Li:2020rqj,Kang:2020fka,H1:2021wkz,Kang:2021kpt,Liu:2021ewb,Kang:2021ffh,Li:2021gjw,Lai:2022aly,Kang:2022dpx,Arratia:2022oxd}. However the intricate jet clustering process usually presents challenges in achieving accurate predictions. Recent advances in this direction can be found in~\cite{Gutierrez-Reyes:2019vbx,Liu:2021xzi}. 
Alternative methods to jets, such as the energy-energy correlator (EEC)~\cite{Basham:1978bw,Basham:1978zq,Hofman:2008ar,Belitsky:2013ofa,Belitsky:2013xxa,Kologlu:2019mfz,Korchemsky:2019nzm,Dixon:2019uzg,Chen:2020vvp} have also been shown to be effective in uncovering the intrinsic transverse dynamics~\cite{Li:2020bub,Ali:2020ksn,Li:2021txc} or the scales of the quark-gluon plasma~\cite{Andres:2022ovj}.

Recently, a novel quantity named the nucleon energy-energy correlator (NEEC) has been proposed as a new look into the nucleon partonic structures~\cite{Liu:2022wop}.  The NEEC probes the initial-final state correlation and takes the form in the momentum fraction $z$ space as~\cite{Liu:2022wop, Liu:2023aqb} 
\bea 
&& f_{q,{\rm EEC}}(z,\theta) \nn \\
&=& \int \frac{dy^-}{4\pi} e^{-izP^+\frac{y^-}{2}} 
\langle P |
{\bar \chi}_n(y^-) \frac{\gamma^+}{2} \hat{{\cal E}}(\theta)\chi_n(0) 
|P \rangle \,, 
\eea 
for the quark NEEC. The gluon NEEC will be given later. Here $\chi_n$ represents the gauge invariant quark field in the Soft Collinear Effective 
Theory (SCET)~\cite{Bauer:2000yr,Bauer:2001yt,Bauer:2001ct,Beneke:2002ph,Bauer:2002nz}.
The definition is equivalent to that of QCD by noting that ${\bar \chi}_n(y^-)\chi_n(0) = {\bar \psi}(y^-) {\cal L}[y,0] \psi(0) $, where ${\cal L}[y^-,0]$ denotes the gauge link between $0$ and $y^-$.
$\hat{{\cal E}}(\theta)$ is the asymptotic energy flow operator~\cite{Sveshnikov:1995vi,Tkachov:1995kk,Korchemsky:1999kt,Bauer:2008dt}, 
that measures the energies from the initial nucleon flowing into the calorimeters sitting far away at angles less than $\theta$. The energy flow at non-zero angles is induced by the intrinsic transverse dynamics. In this sense, studying the $\theta$ distribution of the NEEC allows us to extract information on the intrinsic transverse dynamics of the nucleon/nucleus. 
The Mellin moment of the NEEC is given by $\int dz z^{N-1} f_{\rm EEC}(z,\theta)$. 
Extension of the NEEC to multiple angular correlators by inserting more $\hat{{\cal E}}$ operators at different angles is also attainable. 

 \begin{figure}[htbp]
  \begin{center}
   \includegraphics[scale=0.255]{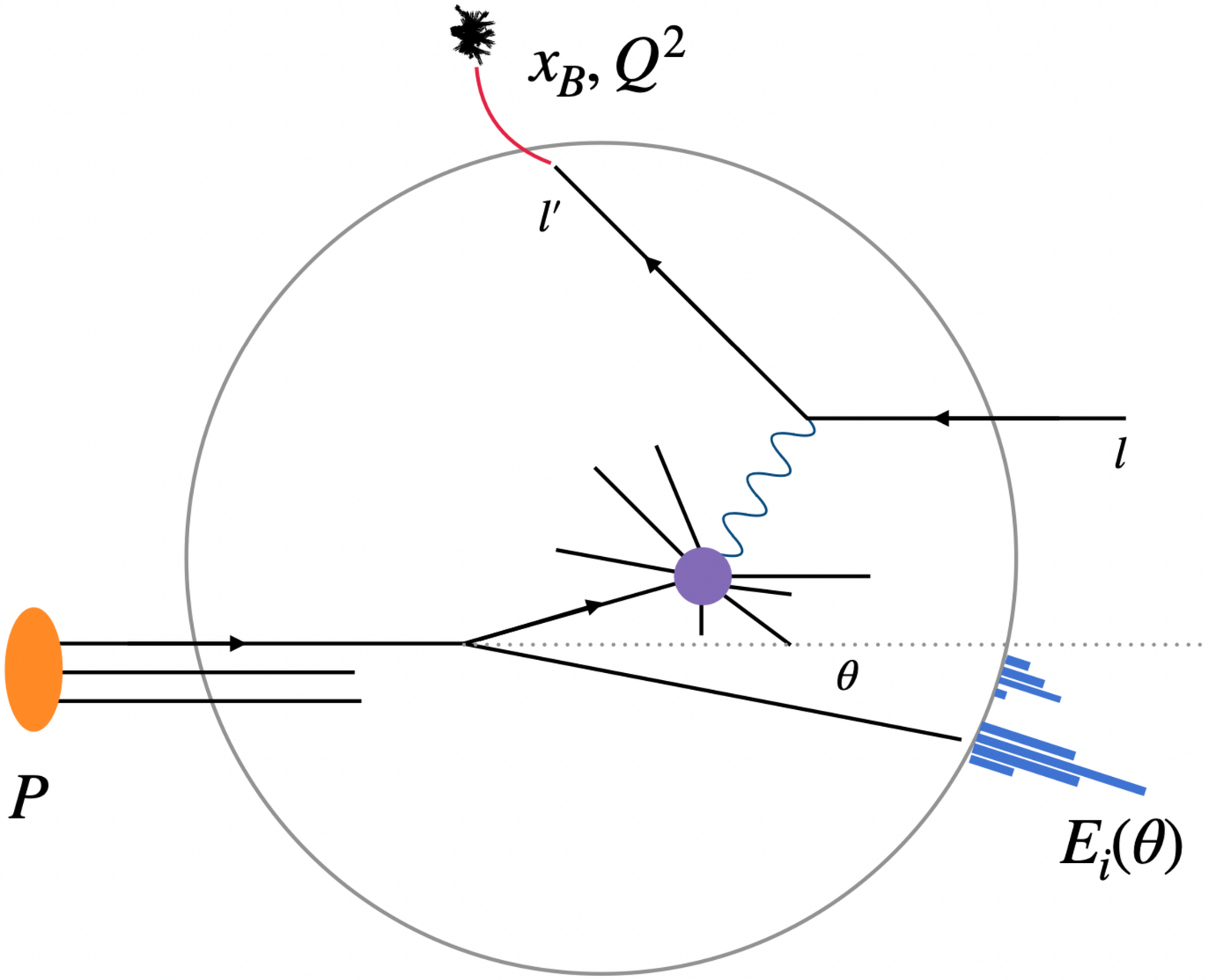} 
\caption{The measurement proposed in Ref.~\cite{Liu:2022wop} as a probe of the NEEC, where the energy $E_i(\theta)$ in the forward detector is recorded. Bjorken-$x_B$ and $Q^2$ are also measured. }
  \label{fg:measure}
 \end{center}
\end{figure}
In Ref.~\cite{Liu:2022wop}, the deep-inelastic scattering (DIS) process illustrated in Fig.~\ref{fg:measure} is suggested to extract the NEEC, in which the energy-weighted cumulant cross section is measured such that
\bea\label{eq:eec-def1} 
\Sigma_N(Q^ 2,\theta) = \sum_i \int d\sigma(x_B,Q^2,p_i) x_B^{N-1} \frac{E_i}{E_P}
\Theta(\theta - \theta_i) \,. 
\eea 
Here $N > 1$ is a positive power and $d\sigma$ is the differential cross section. $x_B$ is the Bjorken variable and $Q^2$ the photon virtuality. The polar angle $\theta_i$ of the calorimetry is measured with respect to the nucleon beam. $p_i$ denotes the momentum flow into the calorimetry and $E_P$ the energy of the incoming nucleon. In this work, we follow Ref.~\cite{Liu:2022wop} to stick the measurement to the Breit frame. We note that experimentally $\Sigma_{N}$ itself is hard to measure, but it is easy to get the measurable $\theta$-distribution by taking the derivative with respect to $\theta$, $d\Sigma_{N}/d\theta$. 

Ref.~\cite{Liu:2022wop} argued without proof, that when $\theta \ll 1$, $\Sigma_N(x_B, Q)$ can be factorized into the partonic DIS cross section ${\hat \sigma}$ and the NEEC to be probed
\bea
\Sigma_N(Q^2,\theta) = \int dx_B x_B^{N-1} \int^1_{x_B}\frac{dz}{z} {\hat \sigma}_i\left(\frac{x_B}{z}\right) f_{i,{\rm EEC}}(z,\theta) \,. 
\eea 
Similar measurement without the $x_B^{N-1}$ weight has also been suggested as a possible access to the gluon saturation phenomena~\cite{Liu:2023aqb} through the $\theta$-distribution of the NEEC.  

However, to reliably extract the NEEC and apply it to the nucleon/nucleus structure studies, the factorization theorem for the $\Sigma_N(Q^2,\theta)$ needs to be reliably established. Meanwhile, sufficient theoretical precision is also required. These serve as the major goals of the current work. In this work, using the SCET framework, we derive the factorization for $\Sigma_N(Q^2,\theta)$. As a first step towards its precision, we carry out the analytic next-to-leading-logarithmic (NLL) resummation for the $\Sigma_{N}(Q^2,\theta)$ when $\theta$ is small and matched onto the ${\cal O}(\alpha_s^2)$ fixed order $\theta$-distribution when $\theta$ becomes large. 

The manuscript is organized as follows. In Section~\ref{sec:factorization}, we show sufficient details on deriving the factorization theorem using SCET. The Section will also present the operator definition of the NEEC $f_{\rm EEC}$. In Section~\ref{sec:pertmatch}, we showed that when $\theta Q \gg \Lambda_{\rm QCD}$, the $f_{\rm EEC}$ can be further matched onto the collinear parton distribution functions (PDFs) with a perturbatively calculable matching coefficient. We discuss its evolution in Section~\ref{sec:evolution}. We calculate all ingredients required for the NLL resummation in Section~\ref{sec:NLLNLO} and predict the small $\theta$-distribution at ${\cal O}(\alpha_s^2)$. The numerical consequence of the resummation and the fixed order $\alpha_s^2$ $\theta$-distribution are studied in Section~\ref{sec:numerics}. We summarize in Section~\ref{sec:summary}.

\section{The Factorization Theorem}\label{sec:factorization}
In this section, we derive the factorization theorem for $\Sigma_N(Q^2,\theta)$ when $\theta Q \ll Q$ using SCET~\cite{Bauer:2000yr,Bauer:2001yt,Bauer:2001ct,Beneke:2002ph,Bauer:2002nz}. Throughout the work, we stick to the Breit frame in which the virtual photon only acquires the momentum in its $z$-component with $q = (0,0,0,-Q)$ and 
proton carries the momentum $P = \frac{Q}{2x_B}(1,0,0,1)$. However, the factorization theorem to be derived is applicable to arbitrary frames. 

The cumulant cross section $\Sigma_N(Q^2,\theta)$ can be calculated by 
\bea\label{eq:def-eef}
&& \Sigma_N(Q^2,\theta) 
=  \frac{\alpha^2 }{Q^4}
\int dx_B x_B^{N-1}
\!\!\! 
\sum_{{\lambda=T,L} }
\!\! e_q^2
f_{\lambda} 
\epsilon^\ast_{\lambda,\mu} \epsilon_{\lambda,\nu} \nn  \\ 
&&   \times
\int 
d^4x
e^{i q \cdot x }
\,  
\langle P |j^{\mu\dagger}(x)    \, 
\hat{{\cal E}}(\theta) \, 
j^\nu  (0) | P \rangle 
\,, 
\eea 
where $e_q$ is the electric charge fraction of the quark initiating the DIS process. Here, we have written the lepton phase space integral as 
$\frac{d^3 l}{(2\pi)^{3} 2 l^0} = 
\frac{Q^2}{16\pi^2 s} dx_B dQ^2
$ and up to vanishing contribution due to the gauge symmetry, we have 
\bea 
&& \sum_{T=1,2}
\epsilon^\ast_{T,\mu} \epsilon_{T,\nu}
= -g_{\mu\nu} + \frac{4x_B^2}{Q^2} P_\mu P_\nu   \,, \nn \\ 
&& 
\epsilon^\ast_{L,\mu} \epsilon_{L,\nu}
= \frac{4x_B^2}{Q^2} P_\mu P_\nu \,, 
\eea 
with $\epsilon^\mu_{T}$ and $\epsilon^\nu_L$ the transverse and longitudinal polarized vector of the virtual photon, respectively. The corresponding flux is given by 
\bea\label{eq:flux} 
f_{T} = 1-y+\frac{y^2}{2}\,, \quad f_L = 2-2y \,, 
\eea 
where $y = \frac{Q^2}{s x_B}$. 
The inserted normalized asymptotic energy flow operator $\hat{{\cal E}}(\theta)$ measures the energy deposited in the detector less than a given angle $\theta$~\cite{Sveshnikov:1995vi,Tkachov:1995kk,Korchemsky:1999kt,Bauer:2008dt} normalized to the energy $E_P$ of the  incoming proton, 
\bea 
\hat{{\cal E}}(\theta) |X\rangle = 
\sum_{i\in X} \frac{E_i}{E_P} \Theta( \theta  - \theta_i ) | X\rangle \,.  
\eea 
We note that if we replace $\hat{{\cal E}}(\theta)$ by the identity operator $1 = \sum_X|X\rangle\langle X|$, Eq.~(\ref{eq:def-eef}) reduces to the definition of the standard DIS cross section.

When $\theta Q \ll Q$, possible leading contribution to the $\Sigma_N(Q^2,\theta)$ comes from the hard degrees of freedom ($H$) whose momentum scale as $p_H=(p_H^+,p_H^-,p_{H,t}) \sim  Q(1,1,1)$~\footnote{Throughout this work, we use the Sudakov decomposition, in which $p^+ = p^0+p^3 \equiv {\bar n}\cdot p$, and $p^- = p^0 - p^3 \equiv n\cdot p$. Here $n=(1,0,0,1)$ and ${\bar n}=(1,0,0,-1)$ while $p_t \cdot n = p_t \cdot {\bar n} =0$.}, the collinear contributions ($C$) with momentum $p_C \sim Q(1,\theta^2,\theta)$, and the soft modes ($S$) with $p_S \sim Q(\theta^a,\theta^a,\theta^a)$ with $a \ge 1$. However, we note that the energy flow operator $\hat{{\cal E}}(\theta)$  acts only on the collinear sector. To see this, we first decompose the final state as $|X \rangle = |X_H X_C X_S \rangle$, and apply the $\hat{{\cal E}}(\theta)$ to find 
\bea\label{eq:HCS} 
\hat{{\cal E}}(\theta) |X\rangle &=&   
\frac{1}{E_P}
\sum_{i\in X} \Big( E_{H,i} \Theta( \theta - \theta_{H,i}) 
+  E_{C,i} \Theta(  \theta - \theta_{C,i} )
\nn \\ 
&&\hspace{5.ex} + E_{S,i} \Theta(  \theta - \theta_{S,i} ) \Big) 
| X_H,X_C,X_S\rangle \,.
\eea 
 Now for the hard radiations, by power counting, $\theta_{H,i} \sim \frac{p_{H,t}}{p_H^+} \sim 1 \gg \theta $ and the $\Theta(\theta  - \theta_{H,i} )$ will hardly be satisfied and therefore the first term in Eq.~(\ref{eq:HCS}) vanishes in the small $\theta$ limit. On the other hand, in the last term, $E_{S,i} \sim \theta^a Q$ is also power suppressed as $\theta \to 0$ when compared with $E_{C,i} \sim Q$. In this way, up to power-suppressed corrections, we find  
\bea 
\hat{{\cal E}}(\theta) |X\rangle &= &
\sum_{i\in X_C} E_{C,i}\Theta(\theta-\theta_{C,i})  
|X_H\rangle |X_C\rangle |X_S\rangle  \nn \\ 
&= & 
(\hat{{\cal E}}(\theta)|X_C\rangle )  |X_H\rangle |X_S \rangle 
\,, 
\eea 
and 
we conclude that
in the small-$\theta$ limit, the measurement $\hat{{\cal E}}(\theta)$ is an operator living solely in the collinear sector and acts inclusively on the hard and the soft radiations. 

To proceed, we follow~\cite{Bauer:2002nz} to match $j^{\mu\dagger}{\cal E}(\theta) j^\nu$ to the SCET operators ${\cal O}_q$ and ${\cal O}_g$, 
with 
\bea 
\langle P| j^{\mu\dagger}(x)  \hat{{\cal E}}(\theta) j^\nu(0) |P \rangle
= C_q^{\mu\nu} \langle P| {\cal O}_{q} |P \rangle  + C_g^{\mu\nu} \langle P|{\cal O}_g |P \rangle \,,
\eea 
where $C_{q/g}^{\mu\nu}$ are the hard matching coefficient to be determined and $C_{g}^{\mu\nu}$ starts at ${\cal O}(\alpha_s)$. The SCET operators are defined as
\bea\label{eq:scetO} 
&&
{\cal O}_q(x,\theta) = 
{\bar \chi}_n(x)Y^\dagger(x) \frac{\gamma^+}{2}  \hat{{\cal E}}(\theta) Y(0)\chi_n(0) \,,\nn \\  
&& 
{\cal O}_g(x,\theta) = 
{\cal B}_\perp (x){\cal Y}^\dagger(x)    \hat{{\cal E}}(\theta) {\cal Y}(0){\cal B}_\perp (0) \,,
\eea 
contains only the gauge invariant collinear quark and gluon fields $\chi$ and ${\cal B}_\perp$, respectively~\cite{Stewart:2010qs}, which are 
\bea 
\chi_n(x) = W_n^\dagger(x) \xi_n(x) \,, \quad 
{\cal B}_{\perp}^\mu 
= \frac{1}{g_s}
[W_n^\dagger i {\cal D}_\perp^\mu W_n](x) \,.
\eea 
We note that both $\chi$ and ${\cal B}_\perp$ scale as $\theta$ by power counting~\cite{Bauer:2000yr}. Here the collinear Wilson lines are
\bea 
W_n(x) 
= \sum_{\rm perms} \exp 
\left(-\frac{g_s}{{\bar n}\cdot P_n} {\bar n}\cdot A_n(x)  \right)  \,, 
\eea 
to make $\chi$ and ${\cal B}_\perp$ gauge invariant. 
 We also have the soft Wilson lines  $Y$ and ${\cal Y}$ in the fundamental and the adjoint representation, respectively. The soft Wilson lines decouple the interaction between the collinear and the soft sectors. Here we note that \bea\label{eq:eycommute} 
[\hat{{\cal E}},Y] = [\hat{{\cal E}},{\cal Y}] = 0\,,
\eea 
since $\hat{{\cal E}}(\theta)$ and $Y({\cal Y})$ act on different sectors. Furthermore, the collinear fields have support in the region where $x^\mu \sim Q^{-1}(1,\theta^{-2},\theta^{-1})$,  while the soft field within the region $x^\mu \sim Q^{-1}(\theta^{-a},\theta^{-a},\theta^{-a})$.  

The hadronic matrix element in Eq.~(\ref{eq:def-eef}) is then matched onto the SCET matrix as 
\bea 
&& \int 
d^4x
e^{i q \cdot x }
\,  
\langle P |j^{\dagger\mu}(x)    \, 
\hat{{\cal E}}(\theta) \, 
j^\nu  (0) | P \rangle  \nn \\ 
&=& \int 
d^4x
e^{i q \cdot x } \nn \\
&& \times 
\Bigg( C_q^{\mu\nu}(x) 
\langle P |
{\bar \chi}_n(x)Y^\dagger(x) \frac{\gamma^+}{2} \hat{{\cal E}}(\theta) Y(0)\chi_n(0)
| P \rangle   \nn \\ 
&& 
+
 C_g^{\mu\nu}(x) 
\langle P| {\cal B}_\perp (x){\cal Y}^\dagger(x)  \hat{{\cal E}}(\theta) {\cal Y}(0){\cal B}_\perp (0)  |P \rangle 
\Bigg)  \,,
\eea 
 where $q \sim Q(1,1,1)$, and hence the $x$ in the hadronic tensor scales as $x \sim \frac{1}{Q}(1,1,1)$. We perform a multiple expansion in the collinear fields and the soft Wilson lines to find 
\bea 
&& \int 
d^4x
e^{i q \cdot x }
\,  
\langle P |j^{\dagger\mu}(x)    \, 
\hat{{\cal E}}(\theta)\, 
j^\nu  (0) | P \rangle  \nn \\ 
&=& \int 
d^4x 
e^{i q \cdot x } \nn \\ 
&& \times 
\Bigg( C_q^{\mu\nu}(x) 
\langle P |
{\bar \chi}_n(x^-)Y^\dagger(0) \frac{\gamma^+}{2} \hat{{\cal E}}(\theta) Y(0)\chi_n(0)
| P \rangle   \nn \\ 
&& +
 C_g^{\mu\nu}(x) 
\langle P| {\cal B}_\perp (x^-){\cal Y}^\dagger(0)  
\hat{{\cal E}}(\theta) {\cal Y}(0){\cal B}_\perp (0)  |P \rangle 
\Bigg)  \,.
\eea 
Now we use the commutation relation between $Y ({\cal Y})$ and ${\cal E}$ in Eq.~(\ref{eq:eycommute}) and the identity $Y^\dagger Y  = {\cal Y}^\dagger {\cal Y}= 1$ to reach
\bea 
&& \int 
d^4x
e^{i q \cdot x }
\,  
\langle P |j^{\dagger\mu}(x)    \, 
\hat{{\cal E}}(\theta) \, 
j^\nu  (0) | P \rangle  \nn \\ 
&=& \int 
d^4x
e^{i q \cdot x }  \Bigg( C_q^{\mu\nu}(x) 
\langle P |
{\bar \chi}_n(x^-) \frac{\gamma^+}{2} \hat{{\cal E}}(\theta)   \chi_n(0)
| P \rangle   \nn \\
&& \hspace{5.ex} +
 C_g^{\mu\nu}(x) 
\langle P| {\cal B}_\perp (x^-)   
\hat{{\cal E}}(\theta)  {\cal B}_\perp (0)  |P \rangle 
\Bigg)  \,.
\eea 
From the derivation, we see clearly that in the small-$\theta$ limit, the measurement is fully inclusive of the soft radiations, and therefore the soft modes do not lead to any logarithmic enhancement contributions. This is different from the conventional TMD measurement, where the soft contribution leads to the enhanced contribution which  eventually gives rise to the perturbative Sudakov factor that suppresses the small transverse momentum region exponentially.

Now we plug the hadronic tensor into Eq.~(\ref{eq:def-eef}) to find the weighted cross section~$\Sigma_N$ takes the form 
\bea 
&& \Sigma_N(Q^2,\theta)
=  \frac{\alpha^2 }{Q^4}
\int dx_B x_B^{N-1}
\!\!\!
\sum_{{\lambda=T,L} }
\!\! e_q^2
f_{\lambda} 
\epsilon^\ast_{\lambda,\mu} \epsilon_{\lambda,\nu} \nn \\ 
&&   \times
\int 
d^4x
e^{i q \cdot x } \Bigg( C_q^{\mu\nu}(x) 
\langle P |
{\bar \chi}_n(x^-) \frac{\gamma^+}{2} \hat{{\cal E}}(\theta)   \chi_n(0)
| P \rangle   \nn \\ 
&& 
\hspace{5.ex} +
 C_g^{\mu\nu}(x) 
\langle P| {\cal B}_\perp (x^-)  \hat{{\cal E}}(\theta)  {\cal B}_\perp (0)  |P \rangle 
\Bigg)
\,.
\eea 

We further manipulate the $\Sigma_N(Q^2,\theta)$ by inserting the complete set $1=|X_C \rangle \langle X_C |$ of the collinear sector into the hadronic tensor, and then perform the translation operation in $x^-$, to find  
\bea 
&& \Sigma_N(Q^2,\theta)
=  \frac{\alpha^2 }{Q^4}
\int dx_B x_B^{N-1}
\!\!\!
\sum_{{\lambda=T,L} }
\!\! e_q^2
f_{\lambda} 
\epsilon^\ast_{\lambda,\mu} \epsilon_{\lambda,\nu} \nn \\ 
&&   \times
P^+ \int  d z \delta((1-z) P^+ - P_C^+)  \int d^4x 
e^{i q \cdot x } 
e^{i(P^+ - P_C^+ ) \frac{x^-}{2}} 
\nn \\ 
&&\times 
\Bigg( C_q^{\mu\nu}(x) 
\langle P |
{\bar \chi}_n(0) \frac{\gamma^+}{2} \hat{{\cal E}}(\theta)   |X_C\rangle \langle X_C| \chi_n(0)
| P \rangle   \nn \\
&& \hspace{5.ex} +
 C_g^{\mu\nu}(x) 
\langle P| {\cal B}_\perp (0)   \hat{{\cal E}}(\theta)  
|X_C\rangle \langle X_C|
{\cal B}_\perp (0)  |P \rangle 
\Bigg)
\,, 
\eea 
where we have inserted the identity  $ P^+ \int dz  \delta((1-z)P^+ - P_C^+)$ to define the variable $z$. Here $P_C^+$ is the large component of the momentum for the collinear radiations. 
Now we replace $(P^+-P_C^+)\frac{x^-}{2}$ in the exponent by $zP^+ \frac{x^-}{2} = z P\cdot x$, where we have used $P \sim P^+ \frac{n^\mu}{2}$ up to ${\cal O}\left(\frac{\Lambda_{\rm QCD}}{Q}\right)$ corrections. With further noticing that $\delta((1-z)P^+ - P_C^+) = \int \frac{dy^-}{4\pi} e^{i[(1-z)P^+-P_C^+]\frac{y^-}{2}}$, and by applying the translation operation on $\langle P| \dots |X_C \rangle $, we find 
the $\Sigma_N(Q^2,\theta)$ possesses the factorized form 
\bea\label{eq:eec-fact1}
  \Sigma_N(Q^2,\theta)
&=&  
\int dx_B x_B^{N-1}
\!\!\!
 \nn \\ 
&&   \hspace{-3.ex}
\times 
 \int  d z \Bigg(   
H_q(z,x_B,Q^2)   \, f_{q,{\rm EEC}}(z,P^+\theta)  \nn \\
&&  +   H_g(z,x_B,Q^2)  \,
f_{g,{\rm EEC}}(z,P^+\theta)
\Bigg)
\,. 
\eea 
where the hard coefficients $H_q$ and $H_g$ are defined as 
\bea 
&&
\hspace{-2.ex}
H_q = \frac{\alpha^2 }{Q^4}
\sum_{{\lambda=T,L} }
 e_q^2
f_{\lambda} 
\epsilon^\ast_{\lambda,\mu} \epsilon_{\lambda,\nu}
 \int 
d^4x
e^{i (q+zP  ) \cdot x } 
 C_q^{\mu\nu}(x) P^+    \,,  \nn \\ 
&& 
\hspace{-2.ex}
H_g = \frac{\alpha^2 }{Q^4}
\sum_{{\lambda=T,L} }
 e_q^2
f_{\lambda} 
\epsilon^\ast_{\lambda,\mu} \epsilon_{\lambda,\nu}
 \int 
d^4x
e^{i (q+zP  ) \cdot x } 
 C_g^{\mu\nu}(x)  \,. 
\eea 
And the collinear functions are 
\bea\label{eq:fqx} 
&& f_{q,{\rm EEC}}(z,P^+\theta) 
=  \int \frac{dy^-}{4\pi} e^{- i z P^+ \frac{y^-}{2} }  \nn \\ 
&&\hspace{10.ex} \times   \langle P |
{\bar \chi}_n\left(\frac{y^-}{2}n^\mu\right) \frac{\gamma^+}{2} 
\hat{{\cal E}}(\theta)  \chi_n(0)
| P \rangle  \,, 
\eea 
for the quarks, and 
\bea\label{eq:fgx}
&& f_{g,{\rm EEC}}(z,P^+\theta)  
= 
\int \frac{dy^-}{4\pi } e^{- i z P^+ \frac{y^-}{2} }   \nn \\ 
&&
\hspace{10.ex} 
\times 
P^+ 
\langle P| {\cal B}_\perp
\left(\frac{y^-}{2}n^\mu \right)   
\hat{{\cal E}}(\theta)   
{\cal B}_\perp (0)  |P \rangle  \,,
\eea 
for the gluon. 
These furnish the operator definition for the quark and gluon {\it nucleon energy-energy correlator} in the momentum space, respectively. 

We can further derive the hard coefficients $H_q$ and $H_g$ by noting that 
\begin{itemize}
\item Once replacing $\hat{{\cal E}}(\theta)$ by the identity operator $1=\sum_X|X\rangle \langle X|$ within the NEEC $f_{i,{\rm EEC}}$, we recover the operator definition for the collinear PDF $f_i(z)$. Meanwhile, Eq.~(\ref{eq:def-eef}) and Eq.~(\ref{eq:eec-fact1}) reduce to the standard inclusive DIS cross section; 
\item The hard coefficients are independent of the details of the collinear sector, and therefore unaffected whether we place the $\hat{{\cal E}}(\theta)$ or the identity operator in the collinear function. 
\end{itemize}
Immediately, we reach the conclusion that the hard functions satisfy 
\bea 
H_q = \frac{1}{z} \, {\hat \sigma}_{q}\left(\frac{x_B}{z},Q^2\right)\,, \quad 
H_g = \frac{1}{z} \, {\hat \sigma}_{g}\left(\frac{x_B}{z},Q^2\right)\,, 
\eea 
and 
are nothing but the DIS partonic cross sections. And therefore
\bea\label{eq:fact-x}
\Sigma_N(Q^2,\theta)
&=&  \sum_{i= q,g} \int dx_B x_B^{N-1}  \nn \\
&& \hspace{-5.ex}
\times \int \frac{dz}{z} \hat{\sigma}_i\left(\frac{x_B}{z},Q^2 \right) f_{i,{\rm EEC}}(z,P^+\theta) \,. 
\eea 
One can observe from the factorization theorem that the $\theta$-dependence of $\Sigma_N(Q^2,\theta)$ is entirely through the $f_{\rm EEC}$, and thus measuring $\Sigma_N(Q^2,\theta)$ directly probes the NEEC. 
The derivation also holds for the measurement without the $x_B^{N-1}$ weighting, as proposed in Ref.~\cite{Liu:2023aqb}, and the factorization is similar to what we have obtained by taking out the integral over $x_B$, which is nothing but the second line of the above equation. 

When $\theta P^+ \gg \Lambda_{\rm QCD}$, as shown in the following section, the NEEC can be matched onto the collinear PDFs, with all $\theta$ dependence occurs only in the perturbative matching coefficients. In this way, since $f_{\rm EEC}$ is dimension-less, the $P^+\theta$ will show up in the form of $\ln\frac{P^+\theta}{\mu}$. Therefore, $\Sigma_{N}$ could also be written as~\footnote{Due to the flux term $f_\lambda(y)$ term from Eq.(\ref{eq:flux}) in the partonic cross section $\hat{\sigma}_i$ with $y= \frac{Q^2}{x_B s} = \frac{Q^2}{s} \frac{1}{u z}$, Eq.~(\ref{eq:fact-N}) should be written as a linear combination of different effective weights $N-i$, for each  $y^i$. However, terms with one power higher in $y$ will be suppressed by $\frac{Q^2}{s}$ for $Q^2\ll s$.  } 
\bea
\label{eq:fact-N}
\Sigma_N(Q^2,\theta)
&=&  \sum_{i= q,g} \int du \, u^{N-1}  \nn \\
&& \hspace{-5.ex}
\times   \hat{\sigma}_i\left(u,Q^2 \right) f_{i,{\rm EEC}}\left(N,\ln \frac{Q\theta}{u\mu}\right) \,, 
\eea 
where $u = \frac{x_B}{z}$ and we have used the fact that $P^+ = \frac{Q}{x_B} =  \frac{Q}{z u}$ in the Breit frame. The $\mu$-dependence in other forms     through the strong coupling and the collinear PDFs are suppressed in the $f_{i,{\rm EEC}}$, where $f_{i,{\rm EEC}}(N,\ln\frac{Q\theta}{u\mu})$ is the NEEC in the Mellin space, 
\bea\label{eq:mellin-s} 
f_{i,{\rm EEC}}(N,\ln\frac{Q\theta}{u \mu})
= \int_0^1 dz\, z^{N-1} f_{i,{\rm EEC}}(z,\ln\frac{ Q\theta}{ z u \mu})\,. \,\, 
\eea 
To simplify the notation, we introduce the $\odot$-product, defined as
\bea\label{eq:odot} 
&& h_1 \odot h_2 \dots \odot h_n \odot f(u)  \nn \\
&=& 
\int \prod_{i}^n du_i u_i^{N-1} h_i(u_i) \,  
f(N,u u_1 u_2\dots u_n)
\eea 
therefore Eq.~(\ref{eq:fact-N}) is written as 
$
\Sigma_N(Q^2,\theta) = 
\hat{\sigma}_i \odot f_{i,{\rm EEC}} (1)\,,
$
we will always drop the ``$(1)$" to write
\bea 
\Sigma_N(Q^2,\theta) = 
\hat{\sigma}_i \odot f_{i,{\rm EEC}}\,.
\eea 

\section{Matching onto the Collinear PDF when $\theta Q \gg \Lambda_{\rm QCD}$}\label{sec:pertmatch}

When $\theta Q \gg \Lambda_{\rm QCD}$, 
the collinear modes can be further split into 
the hard collinear fields $(C_1)$ with momentum scaling $p_{C_1} \sim Q(1,\theta^2,\theta)$ and the $C_2$ modes in SECT$_{\rm II}$ with $p_{C_2} \sim Q(1,\lambda^2,\lambda)$ with $\lambda \equiv \frac{\Lambda_{\rm QCD}}{Q} \ll \theta$. The SCET operators in Eq.~(\ref{eq:scetO}) can be further matched onto the SCET$_{\rm II}$ operators such that 
\bea\label{eq:ItoII} 
{\cal O}_{i}(x^-) = \sum_{j=q,g}
C_{j}(x^-)
{\cal O}_{j,{\rm II}}(x^-) \,,
\eea 
where the operators on the left-hand side of the equation are those that appeared in matrix elements of Eq.~(\ref{eq:fqx}) and~(\ref{eq:fgx}), and the ${\cal O}_{i,{\rm II}}$ is the SCET$_{\rm II}$ operators which have the exact same form as ${\cal O}_i$ but without the energy operator $\hat{{\cal E}}(\theta)$ and is made out of the $C_2$ fields. 

We pause here to study first the effects when $\hat{\cal E}(\theta)$ is acting on $|X\rangle = |X_{C_1},X_{C_2}\rangle $. From the definition, we have 
\bea
&&\hat{\cal E}(\theta)|X_{C_1},X_{C_2} \rangle 
\nn \\ 
&=&
\sum_{
\substack{i\in X_{C_1}\\
j\in X_{C_2} }}
\left( \frac{E_i}{E_P}\Theta(\theta-\theta_i ) 
+ \frac{E_j}{E_P}  \Theta(\theta-\theta_j ) 
\right)
|X \rangle  \,.  
\eea 
We note that since by definition, for particles from the ${C_2}$ modes, the polar angle scales as, $\theta_j \sim \lambda \ll \theta$, the second $\Theta$ function will be always satisfied and can be replaced by $1$. Therefore we find
\bea\label{eq:c1c2} 
&&\hat{\cal E}(\theta)|X_{C_1},X_{C_2} \rangle 
\nn \\ 
&=&\left(
\sum_{i\in X_{C_1}}
 - \frac{E_i}{E_P}\Theta(\theta_i-\theta ) 
+ \frac{E_{X}}{E_P}   
\right)
|X_{C_1},X_{C_2} \rangle \,\, \,   \,.  
\eea 
where $E_{X} \equiv \sum_{i\in X_{C_1},j\in X_{C_2}} E_i+E_j $ and we have used $\Theta(\theta-\theta_i) = 1-\Theta(\theta_i-\theta)$. 

The $E_X$ term in Eq.~(\ref{eq:c1c2}) acts on both $C_1$ and $C_2$ modes simultaneously and contributes to the $f_{\rm EEC}$ in the way that
\bea 
&& f_{i,{\rm EEC}} 
\supset
\sum_{X}
 \frac{E_X}{E_P} 
\langle P|  {\cal O}_i
|X \rangle  
\langle X |{\cal O}_i| P \rangle
\delta((1-z)P^+-P_X^+) 
\nn \\
&& =  
(1-z)
\sum_{ X } 
\langle P|  {\cal O}_i
| X \rangle  
\langle  X |{\cal O}_i| P \rangle \delta((1-z)P^+-P_{X}^+) \nn \\
&& = f_i(z) - z\, f_i(z) \,, \qquad \text{with $i=q,g$\,,}
\eea 
where $f_i(z)$ is the collinear PDF. In the first line we have inserted the complete set $\sum_X |X\rangle \langle X|$ into Eq.~(\ref{eq:fqx}) and~(\ref{eq:fgx}) and applied Eq.~(\ref{eq:c1c2}) but only kept the $E_X$ term. We performed the translation operation in the $n^\mu$ direction before we integrate over $y^-$. Here, we have also used the definition of the collinear PDF
\bea\label{eq:pdf} 
 f_i(z) 
&=&  \sum_X \langle P |{\cal O}_i |X \rangle 
\langle X| {\cal O}_i | P \rangle 
\delta((1-z)P^+ - P_X^+) \nn \\ 
&=&  \int \frac{dy^-}{4\pi}
e^{-izP^+\frac{y^-}{2}}
\langle P|{\cal O}_i\left(y^-\frac{n^\mu}{2}\right)
{\cal O}_i |P \rangle \,. 
\eea 

The $- E_i\theta(\theta_i-\theta)$ term in Eq.~(\ref{eq:c1c2}) acts only on the $C_1$ modes. Therefore when matching onto SCET$_{\rm II}$, toghter with the coefficient $C_j(x^-)$ in Eq.~(\ref{eq:ItoII}), it gives the matching coefficient. The matching procedure is similar to what we did in the previous section and we will not repeat it here. The final contribution from the $- E_i\theta(\theta_i-\theta)$ term then reads
\bea 
f_{i,{\rm EEC}} \supset - \sum_j \int^1_z \frac{d\xi}{\xi} I'_{ij}\left(\frac{z}{\xi} ,\ln\frac{z Q\theta}{x_B \mu}\right)\,  \left[ \xi f_j(\xi) \right] \,, 
\eea 
where the additional $\xi$ in front of $f(\xi)$ originates from $E_i/E_P$. Here $I_{ij}'$ is the matching coefficient that can be calculated perturbatively and starts from ${\cal O}(\alpha_s)$.  

Gathering all pieces, the matching of the NEEC $f_{i,{\rm EEC}}$ to the collinear PDFs when $\theta Q \gg \Lambda_{\rm QCD}$ is given by

\bea\label{eq:Ix} 
&&f_{i,{\rm EEC}}(z,\ln\frac{Q \theta}{u \mu}) \nn \\ 
&=& f_i(z) - \int_z^1 \frac{d \xi}{\xi} 
I_{ij}\left(
\frac{z}{\xi}, \ln\frac{Q\theta}{u\mu} 
\right) \,  \xi f_j(\xi) \,, 
\eea 
where $I_{ij}(z) = \delta(1-z) + I'_{ij}(z)$. It will be interesting to note that the $\theta$ dependence is solely within the $\xi f_j(\xi)$ term where $I_{ij}$, as we will show, is determined by the splitting function $P(z_i,\dots)$, involves the factor $z_i P(z_i,\dots)$. Here, the $z_i$ factor originated from the energy weight of parton $i$. Therefore, from Eq.~(\ref{eq:Ix}), we can interpret $df_{\rm EEC}/d\theta$ as the parton energy density at the angle $\theta$ for the given incoming parton energy density $\xi f(\xi)$.  

Written in the Mellin space, we have 
\bea\label{eq:IN} 
&&f_{i,{\rm EEC}}(N,\ln\frac{Q\theta}{u \mu}) \nn \\ 
&=& f_i(N) - 
I_{ij}\left(
N, \ln\frac{Q\theta}{u \mu}
\right) f_j(N+1) \,, 
\eea 
where $I_{ij}(N)$ is the Mellin moment of $I_{ij}(z)$. 

For later use, we define the $\ast$-product by
\bea\label{eq:astdot} 
I\ast f = I(N)f(N+1) \,. 
\eea 
With this notation, Eq.~(\ref{eq:IN}) is written as 
\bea 
f_{i,{\rm EEC}} = f_i - \sum_j I_{ij}\ast f_j \,. 
\eea 
We note the difference between $f_i = f_i(N)$ and $1\ast f_i = f_i(N+1)$.

\section{Evolution Equations} \label{sec:evolution}
From the factorization theorem in Eq.~(\ref{eq:fact-x}), Eq.~(\ref{eq:fact-N}), 
and the consistency relation  
\bea 
\frac{d}{d\ln\mu^2} \Sigma_N(Q^2,\theta) = 0  \,, 
\eea 
we deduce that the NEEC satisfies the \emph{modified} Dokshitzer–Gribov–Lipatov–Altarelli–Parisi (DGLAP) evolution equation
\bea\label{eq:evo1} 
&& \frac{d}{d\ln\mu^2} f_{i,{\rm EEC}}(z,\ln\frac{ Q\theta}{z u\mu}) \nn \\ 
&=& \sum_j \int_z^1 \frac{d\xi}{\xi}  P_{ij}\left(\frac{z}{\xi}\right)
 f_{j,{\rm EEC}}(\xi,
 \ln \frac{Q\theta}{ z u \mu })\,,
\eea 
in the momentum space. The inclusion of $z$, as an argument of the function $f_{j, \rm EEC}$ indicates that \eqref{eq:evo1} cannot be considered a conventional convolution beyond LL accuracy. The presence of this extra dependency arises from the inherent angular nature of NEEC, which results in its reliance on the frame of reference in which the observation is made. An analogous situation is observed in the case of the final state EEC, as discussed in \cite{Dixon:2019uzg}.
In the Mellin space, the evolution of the NEEC follows
\bea \label{eq:evo2}
&& \frac{d}{d\ln\mu^2} f_{i,{\rm EEC}}(N,\ln\frac{Q\theta}{u \mu}) \nn \\
&=&
\sum_j \int d\xi \xi^{N-1} 
P_{ij}\left(\xi\right)
 f_{j,{\rm EEC}}(N,\ln\frac{Q\theta}{\xi \, u \mu}) \nn \\ 
 &=& 
 P_{ij}\odot f_{j,{\rm EEC}}(u)
 \,, 
\eea 
where $P_{ij}(\xi)$ is the vacuum splitting function and the $\odot$ notation follows Eq.~(\ref{eq:odot}). Note that the additional $\xi$ within the logarithm is due to the specific structure of the Mellin transformation for $f_{\rm EEC}$ in Eq.~(\ref{eq:mellin-s}). 

In the momentum space, the solution to the evolution equation Eq.~(\ref{eq:evo2}) can be solved numerically using {\tt HOPPET}~\cite{Salam:2008qg} or {\tt APFEL++}~\cite{Bertone:2013vaa} with the initial condition at $\mu_0 \sim Q\theta$ to be determined later in Section~\ref{sec:collinear}. The solution in the Mellin space is slightly more involved and we solve it in Appendix~\ref{sec:derive}. Its analytic form will be given in Section~\ref{sec:nll}.  

In practice, it is useful to introduce for the NEEC $f_{{\rm EEC}}$ the flavor singlet and non-singlet distributions, where the singlet part is given by 
\bea 
F^{S}_{q}  = \sum_i (f_{q_i,{\rm EEC}} 
+ f_{{\bar q}_i,{\rm EEC}}) \,, \quad 
F^{S}_g = f_{g,{\rm EEC}} \,,
\eea 
and the non-singlet part is defined as 
\bea 
F^{NS}_{i} 
= N_F(f_{q_i,{\rm EEC}} 
+ f_{{\bar q}_i,{\rm EEC}})
- F^{S}_{q} \,. 
\eea 
The definition follows directly those of the collinear PDFs~\cite{Moch:2004pa,Vogt:2004mw}. We note that
\bea
f_{q_i,{\rm EEC}} + f_{{\bar q}_i,{\rm EEC}}
= \frac{1}{N_F}  (F_{i}^{NS} + F_q^{S}) \,, \,\,
f_{g,{\rm EEC}} = F^{S}_g \,. 
\eea 
Since the $f_{\rm EEC}$ behaves exactly like the collinear PDF, by the flavor and charge conjugation symmetry, 
the non-singlet distribution for the NEEC evolves as~\cite{Moch:2004pa,Vogt:2004mw}
\bea\label{eq:evos}  
\frac{d}{d\ln\mu^2} F_{i}^{NS}(N,\ln\frac{Q\theta}{u\mu})
=  P_{NS}^+ \odot F_{i}^{NS}(u)
\,, 
\eea 
with no mixing with the singlet distributions $F^{S}_{q}$ and $F^{S}_{g}$.
Here `$\odot$' follows Eq.~(\ref{eq:odot}). The singlet distributions evolve as  
\bea\label{eq:evons}
 \frac{d}{d\ln\mu^2}\begin{bmatrix}
F_{q}^{S}    \\ 
F_{g}^{S}    
\end{bmatrix} = 
\begin{bmatrix}
P^{S}_{qq} & P^{S}_{qg}   \\ 
P^{S}_{gq} & P^{S}_{gg}   \\
\end{bmatrix} 
\odot 
\begin{bmatrix}
F_{q}^{S}    \\ 
F_{g}^{S}    
\end{bmatrix}(u)  \,.
\eea 
Here $P^{S}_{gg} = P_{gg}$, $P^{S}_{qg} = 2N_F P_{qg}$,  $P^{S}_{gq} =  P_{gq}$ and 
\bea 
P_{qq}^{S} = P_{NS}^+ + P_{ps}\,,
\eea 
are defined in Ref.~\cite{Moch:2004pa,Vogt:2004mw}. The non-singlet and the pure singlet splitting kernels $P_{NS}^+$
and $P_{ps}$ can also be found therein. At order $\alpha_s$, $P_{NS}^+ = P_{qq}$ and $ P_{ps} = 0$.
The ${\cal O}(\alpha_s^2)$ results are given in the Appendix.

The evolution of the matching coefficient $I_{ij}$ can be directly derived from Eq.~(\ref{eq:Ix}) and Eq.~(\ref{eq:IN}) along with the evolution of the $f_{\rm EEC}$ in Eq.~(\ref{eq:evo1}) to Eq.~(\ref{eq:evo2}). For practical use, we note that by the charge conjugation and flavor symmetry, the matching coefficient $I_{ij}$ for the quark can always be written as 
\bea 
&& I_{q_iq_j} = I_{{\bar q}_i{\bar q}_j} = I_{qq}^{NS}\delta_{ij} + I^{PS}_{qq} \,, \nn \\ 
&& I_{q_i{\bar q}_j} = I_{{\bar q}_iq_j} = I_{q{\bar q}}^{NS}\delta_{ij} + I^{PS}_{q{\bar q}}\,, 
\eea 
where $I^{PS}$ is flavor independent. In this way, we find 
\bea 
F_{i}^{NS} = 
f_{i}^{NS} - 
(I_{qq}^{NS}+I_{q{\bar q}}^{NS}) \ast f_{i}^{NS}
\equiv f_{i}^{NS} - I^{NS} \ast f_{i}^{NS} 
\,, 
\eea 
and the {\it pure} quark contribution to $F_q^S$ is 
\bea 
F_{q}^{S} &=& f_{q}^{S} - (I_{qq}^{NS}+I_{q{\bar q}}^{NS}+N_F(I_{qq}^{PS}+I_{q{\bar q}}^{PS}))   \ast f_{q}^{S} \nn \\
&\equiv& f_{q}^{S} - I_{qq}^{S} \ast f_{q}^{S} \,, 
\eea 
where we follow Eq.~(\ref{eq:astdot}) to use `$\ast$' as the shorthand notation for the product in Eq.~(\ref{eq:IN}). Up to order $\alpha_s$, $I_{qq}^{PS}=I_{q{\bar q}}^{PS} = I_{q{\bar q}}^{NS} = 0$, and thus $I^{NS}  = I_{qq}^S = I_{qq}^{NS}$ up to this order. 
Here 
\bea 
f^{NS}_{i} = N_F(f_{q_i} + f_{{\bar q}_i}) -\sum_k (f_{q_k} + f_{{\bar q}_k}) \,,
\eea 
is the singlet PDF distribution 
and 
\bea 
f^{S}_{q} = \sum_i f_{q_i} + f_{{\bar q}_i}\,,\quad 
f^{S}_g = f_g\,,
\eea 
are the non-singlet distributions. They satisfy the same DGLAP evolution in Eq.~(\ref{eq:evos}) and Eq.~(\ref{eq:evons}) after replacing $\odot$ by the product, for the singlet and the non-singlet PDFs, respectively. 

It is immediately realized that 
\bea\label{eq:evo3} 
\frac{d}{d\ln\mu^2} I^{NS}(N,u)
 = P_{NS}^+ \odot I^{NS}(u) 
 - I^{NS} \ast P_{NS}^+ \,, 
 \eea 
 and 
 \bea \label{eq:evo4} 
\frac{d}{d\ln\mu^2} I_{ij}^{S}(N,u)
 = P_{ik}^{S}  \odot I_{kj}^{S}(u) 
 - I_{ik}^{S} \ast P_{kj}^{S}  \,. 
 \eea 
Here $i = q,g$. Here summation over the repeat indices is assumed. 


\section{Matching coefficients at NLO and the NLL resummation}\label{sec:NLLNLO}
In this section, we calculate the cumulant cross section 
\bea 
\Sigma_N(Q^2,\theta)
&=&  \sum_{i= q,g} \hat{ \sigma_i } \odot f_{i,{\rm EEC}}  \,. 
\eea 
to NLO in $\alpha_s$ in the small $\theta$ limit. The $\odot$-product follows Eq.~(\ref{eq:odot}). The ${\cal O}(\alpha_s)$ calculation allows us to realize the NLL resummation for $\Sigma_N(Q^2,\theta)$, which in turn will allow us to predict the complete $\alpha_s^2$ distribution $\frac{d}{d\theta^2} \Sigma_N(Q^2,\theta)$ when $\theta$ is small. 

  \subsection{NLO Hard Function}\label{sec:hard}
  For the NLL resummation, we need the DIS partonic cross section at NLO. The NLO partonic cross section $\hat{\sigma}(z,Q^2)$ is well-known~\cite{Bardeen:1978yd,Altarelli:1978id,Humpert:1980uv} and we present the results in the Appendix. Here we supply the cross section in the Mellin space, which can be written as 
\bea\label{eq:sighat} 
\hat{\sigma}(N,Q^2) 
=\frac{4\pi\alpha^2}{Q^4}  \sum_{i=-N_F}^{N_F}
\sum_{c=q,g}\sum_{\lambda = T,L} e^2_{q_i} f_\lambda \,  \hat{\sigma}_{c,\lambda}(N)\,,
\eea 
where
\bea 
\hat{\sigma}_{c,\lambda}(N) = \sum_{n=0} \left(\frac{\alpha_s}{2\pi}\right)^n \hat{\sigma}^{(n)}_{c,\lambda} (N) \,,
\eea 
in which at LO
\bea 
\hat{\sigma}^{(0)}_{q,T} = 1 \,, \quad 
\hat{\sigma}^{(0)}_{q,L} = \hat{\sigma}^{(0)}_{g,T} =\hat{\sigma}^{(0)}_{g,L} = 0 \,.  
\eea 

To obtain compact results for $\sigma_{c,\lambda}^{(1)}(N)$ at NLO, we introduce the $S_{\pm m}$ and $S_{\pm m_1,m_2,\dots}$ functions~\cite{Moch:2004pa,Vogt:2004mw} defined as 
\bea\label{eq:S1} 
S_{\pm m }(N) = \sum_{i=1}^N
\frac{(\pm 1)^i}{i^m} \,, 
\eea 
and 
\bea \label{eq:S2}
S_{\pm m_1,m_2,\dots,m_k } (N) = \sum_{i=1}^N
\frac{(\pm 1)^i}{i^{m_1}}
S_{m_2,\dots,m_k }(i) \,,
\eea 
and we introduce $N_{\pm k} S_{\vec{m}}(N) = S_{\vec{m}}(N\pm k)$ raise/lower the argument by $k$. We abbreviates $S_{\vec{m}}(N) = S_{\vec{m}}$. 
Some useful formulae are presented in the Appendix. 

We thus find 
the quark contribution to $\hat{\sigma}^{(1)}$ reads 
\bea
 \hat{\sigma}^{(1)}_{q,L}
=  C_F (-)(1-\hat{N}_+)S_1  \,, 
\eea 
for the longitudinal part, where 
 we have used $1 = (1-z)\sum_{i=0}^\infty z^i $ anywhere necessary to get the results, 
and 
\bea
&&  \hat{\sigma}^{(1)}_{q,T} =  C_F  \Bigg\{ 
\left( \frac{3}{2} - (\hat{N}_- + \hat{N}_+)S_1 \right)\ln\frac{Q^2}{\mu^2}
\nn \\ 
&&+ (\hat{N}_- + \hat{N}_+)(S_{1,1}-S_2) + \frac{\pi^2}{3} \nn \\ 
&& + \frac{3}{2}(\hat{N}_-)S_1
-3(\hat{N}_- - 1)S_1
  - \left( \frac{9}{2} + \frac{\pi^2}{3} \right) 
\Bigg\} 
\eea 

The gluon channel is given by 
\bea 
\hat{\sigma}_{L,g}^{(1)} =   T_R 
(-2)(1-2\hat{N}_++\hat{N}_{+2}) S_1\,, 
\eea 
and 
\bea 
&& \hat{\sigma}^{(1)}_{T,g} =  \, 
T_R \Bigg\{(- \hat{N}_- + 3 - 4\hat{N}_+ + 2 \hat{N}_{+2})
S_1 \ln\frac{Q^2}{\mu^2} \nn \\ 
&&
+(\hat{N}_- 
- 3 + 4\hat{N}_+ - 2\hat{N}_{+2})(S_{1,1}-S_2) 
\nn \\ 
&&
+(-5+\hat{N}_- + 8\hat{N}_+ - 4\hat{N}_{+2})S_1
\Bigg \} \,, 
\eea 
where a factor of $1/2$ has been multiplied into the gluon channel to match with the flavor summation in Eq.~(\ref{eq:sighat}). 

\subsection{NLO Matching Coefficient for $I_{ij}$}\label{sec:collinear}
The matching coefficients $I_{ij}$ in Eq.~(\ref{eq:IN}) can be obtained by calculating the difference between the NEEC defined in Eq.~(\ref{eq:fqx}), (\ref{eq:fgx}) and the collinear PDF in Eq.~(\ref{eq:pdf}), using the SCET Feynman rules. To perform the matching, the external hadronic states $|P\rangle$ and $|X\rangle$ can be replaced by the partonic degrees of freedom, using on-shell quarks and gluons. In dimensional regularization, the higher-order corrections to Eq.~(\ref{eq:pdf}) are dimensionless and vanish identically. Therefore, the $I_{ij}$ is determined by calculating the matrix elements in 
 Eq.~(\ref{eq:fqx}) and (\ref{eq:fgx}). At NLO, this results in evaluating the phase space integrals of the form
\bea 
&&f_{i,{\rm EEC}} 
= - P^+ \int dz z^{N-1} \int d\xi  \, 
\delta((\xi - z)P^+ - g^+) 
\nn \\ 
&\times & \int \frac{d^d g}{(2\pi)^{d-1}}
\delta(g^2) \, 
\left(1-\frac{z}{\xi} \right) \xi \,
\Theta(\theta_g - \theta) \, 
\nn \\ 
&\times &
(8\pi\alpha_s)  \mu^{2\epsilon}
\frac{1-\frac{z}{\xi}}{g_t^2} P^{(0)}_{ij}
\left( \frac{z}{\xi},\epsilon \right)
f_j(\xi)   \,, 
\eea 
where $g^\mu$ is the momentum of the detected parton, and $g_t$ is its transverse component. $\xi$ is the momentum fraction carried by the incoming parton. 
Here $P^{(0)}_{ij}$ are the ${\cal O}(\alpha_s)$ splitting kernels, which are
\bea\label{eq:split}
&& P^{(0)}_{qq}(z,\epsilon) = C_F \left( \frac{1+z^2}{1-z}
-\epsilon(1-z) \right) \,, \nn \\ 
&& P^{(0)}_{gq}(z,\epsilon) = 
C_F
\left( \frac{1+(1-z)^2}{z} -\epsilon z
\right) \,, \nn \\ 
&& P^{(0)}_{qg}(z,\epsilon) = 
T_R
\left( z^2+(1-z)^2 - 2\epsilon z(1-z)  
\right) \,, \nn \\ 
&& P^{(0)}_{gg}(z,\epsilon) =  
2C_A\left(
\frac{z}{1-z}
+ \frac{1-z}{z}
+z(1-z) 
\right)
\eea

To evaluate the integral, we parameterize the phase space as 
\bea 
 \frac{d^dg}{(2\pi)^{d-1}}\delta(g^2) 
= \frac{1}{16 \pi^2} 
\frac{(4\pi)^\epsilon}{\Gamma(1-\epsilon)}
\frac{dg^+}{g^+}
( \frac{g^+}{2})^{2-2\epsilon}
d\theta_g^2  
\theta_g^{-2\epsilon} 
\,, 
\eea 
where we have used $g_t = \theta_g \frac{g^+}{2}$. We thus find the NLO result of Eq.~(\ref{eq:IN}) is 
\bea 
\hspace{-3.ex}
f_{i,{\rm EEC}} 
= f_{i}(N) - \left(\delta_{ij}
+ \frac{\alpha_s}{2\pi} \, I^{(1)}_{ij}(N) \right) 
f_j(N+1)   \,,  
\eea 
where the un-renormalized NLO matching coefficient is
\bea 
I_{ij}^{(1)} &=& \frac{ 1}{ \epsilon } 
\frac{1}{\Gamma(1-\epsilon)}
\left( \frac{4\pi \mu^2}{(\frac{Q\theta}{2u})^2} \right)^\epsilon \nn \\ 
&\times & 
\int dz z^{N-1} \, 
\left( \frac{z}{1 - z} \right)^{2\epsilon}
\left(1-z \right) 
P^{(0)}_{ij}
\left( z,\epsilon \right) \,. 
\eea 

Plugging the splitting functions in Eq.~(\ref{eq:split}), we find the NLO unrenormalized matching coefficients
\bea 
I_{ij,un}^{(1)}
&=&
S_\epsilon
\left(\frac{1}{\epsilon} 
+  \ln \frac{ \mu^2}{\frac{Q^2}{4u^2}\theta^2}
\right)\left[ 
P^{(0)}_{ij}(N) - P^{(0)}_{ij}(N+1)
\right] \nn \\ 
&+ & 
 d^{(1)}_{ij}(N)
- d^{(1)}_{ij}(N+1) \,, 
\eea 
where the angular factor 
$S_\epsilon = \frac{(4\pi)^\epsilon}{\Gamma(1-\epsilon)}$. 
$P_{ij}^{(0)}(N)$ are the ${\cal O}(\alpha_s)$ splitting functions in Mellin space, which are 
\bea\label{eq:PN0} 
 && P_{qq}^{(0)}(N) 
= 
 C_F  \left( \frac{3}{2} -  (\hat{N}_++\hat{N}_-)S_1 \right) 
\,,    \nn \\ 
&& P_{gq}^{(0)}(N) 
=
 C_F 
(-2\hat{N}_{-2}+4\hat{N}_{-}+\hat{N}_+ -3) S_1 
\,,  \nn \\ 
&& P_{qg}^{(0)}(N) 
=  T_R 
(-\hat{N}_{-}-4\hat{N}_{+}+2\hat{N}_{+2}+3) S_1 
\,,   \nn  \\ 
&& P_{gg}^{(0)}(N) \nn \\ 
&=&
 2C_A \left[
-\hat{N}_{-2}+2(\hat{N}_{-}+\hat{N}_{+})-\hat{N}_{+2}-3\right]  S_1 
+ \frac{\beta_0}{2}
\,,    \nn \\ 
\eea 
where $\beta_0 = \frac{11}{3}C_A - \frac{4}{3}N_F T_R$. The splitting functions in the $z$-space are well-known and can be found in the Appendix. 

The NLO $\theta$ independent constant terms are calculated to find the general form 
\bea 
d_{ij}^{(1)}(z) 
=    2p^{(0),0}_{ij}(z)
\ln \frac{z}{1-z} + p^{(0),1}_{ij}(z)  \,, 
\eea 
and 
\bea 
d_{ij}^{(1)}(N) = \int dz z^{N-1} d_{ij}^{(1)}(z) \,,  
\eea 
where $p_{ij}^{(0),k}(z)$ are the coefficients of the $\epsilon^k$ with $k=0,1$ in the splitting kernels $P_{ij}^{(0)}(z,\epsilon)$ of Eq.~(\ref{eq:split}). Here All divergences for $z \to 1$ are understood in the sense of $+$-distributions.

Evaluating the Mellin integral, we find 
\bea 
 d^{(1)}_{qq}(N)   
&=&
2 C_F\left[(\hat{N}_++\hat{N}_-)(S_2-S_{1,1}) 
-\frac{\pi^2}{3} 
\right. \nn \\
&& \left. 
+ (\hat{N}_++\hat{N}_- -2)\frac{S_1}{2} \right] \,,  \nn 
\\ 
d^{(1)}_{gq}(N)  &=& 
2 C_F\Bigg[ (2\hat{N}_{-2}-4\hat{N}_{-}+3-\hat{N}_{+} )(S_2-S_{1,1})  \nn \\
&& 
+(1-\hat{N}_+)\frac{S_1}{2}
\Bigg] \,,  \nn 
\\ 
d^{(1)}_{qg}(N)   
&=&
2T_R \Bigg[ (\hat{N}_- - 3 
+4 \hat{N}_+ 
-2 \hat{N}_{+2})(S_2-S_{1,1}) \nn \\
&& 
+(1-2\hat{N}_++\hat{N}_{+2})S_1
\Bigg]\,, \nn \\
 d_{gg}^{(1)}(N)   
&=& 4C_A\left[ \left(
3-2(\hat{N}_+ + \hat{N}_-)
\right. \right. 
\nn \\ 
&& \left.\left. 
+(\hat{N}_{+2} + \hat{N}_{-2})
\right )(S_2-S_{1,1})
-\frac{\pi^2}{6}
\right] \,. 
\eea 
The NLO  renormalized matching coefficient in Eq.~(\ref{eq:IN}) is then 
\bea\label{eq:Iij} 
I_{ij} 
&=& \delta_{ij} 
+ \frac{\alpha_s}{2\pi} \left[ - 
\ln\frac{Q\theta}{2u\mu}\Big(2 P_{ij}^{(0)}(N)-2P_{ij}^{(0)} (N+1) \Big) \right. \nn \\
&&
\left.
\hspace{10.ex}+ 
  d_{ij}^{(1)}(N)-d_{ij}^{(1)}(N+1) \right] \,. 
\eea 
The NLO calculation explicitly verified the evolution equation derived via the consistency condition in Section~\ref{sec:evolution}. From the calculation, we can also read the singlet and the non-singlet terms introduced in Section~\ref{sec:evolution}, which are 
\bea 
&& I^{NS} = I_{qq}^S =  I_{qq}\,, \qquad
I^{S}_{gg} = I_{gg} \,, \nn \\ 
&& I^{S}_{qg} = 2 N_F I_{qg} \,, \qquad 
I^{S}_{qg} =   I_{qg} \,. 
\eea 

\subsection{NLL Resummation for $\Sigma_N$}\label{sec:nll}
When $\alpha_s \ln \theta^2 \sim 1$, the logarithmic terms are large and should be resummed to all orders, the NLO calculations in the previous section allow us to realize the NLL resummation for the NEEC, namely the resummation of $\alpha_s^k \ln^{k}\theta^2$ and $\alpha_s^k \ln^{k-1}\theta^2$ series.

One way to perform the resummation is to evaluate the partonic cross section $\hat{\sigma}_{c,\lambda}$ and the PDFs $f_i$ at scale $\mu \sim Q$, and evolve the matching coefficient $I$ from $\mu_0 \sim Q\theta$ to $\mu \sim Q$ following the resummation equation in Eq.~(\ref{eq:evo3}) and Eq.~(\ref{eq:evo4}) in Section~\ref{sec:evolution}. 

Equivalently, we can also set the scales for both the collinear PDFs $f^{NS}_{q_i}$, $f_j^S$, and the matching coefficients $I^{NS}$, $I_{ij}^S$ at $\mu_0 \sim Q\theta$, to evaluate the NEEC and evolve the NEEC from $\mu_0$ to $\mu$ to realize the resummation. In the $z$-space, the evolution is identical to the collinear PDFs and can be achieved numerically by  {\tt HOPPET}~\cite{Salam:2008qg} or {\tt APFEL++}~\cite{Bertone:2013vaa}. In the Mellin space, 
the resummation follows the evolution equations in Eq.~(\ref{eq:evos}) and (\ref{eq:evons}) for both $F_{q_i}^{NS}$ and $F^S_i$.  
%
We solve the equations iteratively in Appendix~\ref{sec:derive}, and find that the NLL NEEC receives the compact analytic form 
\bea\label{eq:fNLLevo} 
&& f_{i,{\rm EEC}}(\mu) = f_i(N,\mu)  -    
   {\cal D}_{ik}^N(\mu,\mu_0)
  \, I_{kj}(u\mu_0) f_{j}(N+1,\mu_0) \nn \\ 
 && - 
 \frac{\alpha_s(\mu_0)}{2\pi}
 {\cal N}_{ik}   [2P_{kj}^{(0)}(N)-2P_{kj}^{(0)}(N+1)]f_j(N+1,\mu_0) \,. \nn \\ 
\eea 
The resummed form holds for both the singlet $F^S$ and non-singlet distributions $F^{NS}$.   
Here $I_{ij}(u\mu_0)$ is the NLO matching coefficient in Eq.~(\ref{eq:Iij}) evaluated at scale $\mu_0$, and the evolution factor ${\cal D}^N_{ij}(\mu,\mu_0)$ is nothing but the DGLAP evolution in the Mellin space, 
\bea\label{eq:D}
{\cal D}^N_{ij}(\mu,\mu_0) = 
\exp\left[\int_{\mu_0}^{\mu} d\ln\mu^2 P(N,\mu)\right]_{ij} \,.
\eea 
To realize the NLL resummation, we need $P_{ij}(N)$ at LO and NLO within the evolution factor ${\cal D}^N_{ij}$. The LO results have been presented in Eq.~(\ref{eq:PN0}), and the NLO moments can be found in Ref.~\cite{Moch:2004pa,Vogt:2004mw} and are also given in the Appendix. Note that we need to divide the $\gamma_{ij}^{(1)}$'s therein by a factor $(-4)$ to get $P_{ij}^{(1)}(N)$ in our normalization. 

The correction to the DGLAP evolution starts from $\alpha_s^nL^{n-1}$ order, in which  
\bea
{\cal N}_{ij}
&=&  \int_{\mu_0}^\mu d\ln{\mu_1^2}
{\cal D}_{ik}^N(\mu,\mu_1) \tilde{P}_{kl}(N,\mu_1)
{\cal D}_{lj}^N(\mu_1,\mu_0) \,, \quad 
\eea 
 originated from the $- \frac{\alpha_s}{2\pi} P \odot \ln u $ where the $\ln u$ term comes from the NLO matching coefficient $I_{ij}$. We note that both ${\cal D}$ and ${\cal N}$ can be integrated analytically using the formula in Eq.~(\ref{eq:ana1}) and Eq.~(\ref{eq:ana2}) of the Appendix. Here we have defined 
\bea\label{eq:Ptilde} 
\tilde{P}_{ij} (N)\equiv 
\int dz z^{N-1} P_{ij}(z) \ln z
= \partial_N P_{ij}(N) \,.
\eea 
Note that derivative of Mellin moment has also appeared in the solution of small angle EEC~\cite{Dixon:2019uzg}.
For the NLL resummation, we need 
\bea\label{eq:Ptildeij} 
&& \tilde{P}_{qq}^{(0)}(N) =C_F \left((\hat{N}_++\hat{N}_-)S_2 -\frac{\pi^2}{3} 
\right) 
\,,  \nn \\ 
&& \tilde{P}_{qg}^{(0)}(N)
= T_R (\hat{N}_- -3 +4 \hat{N}_+ - 2 \hat{N}_{+2})S_2 \,, \nn \\ 
&& \tilde{P}_{gq}^{(0)}(N) = 
C_F\left(2\hat{N}_{-2}
-4\hat{N}_- + 3  -  \hat{N}_+
\right)S_2
\,, \nn \\
&&\tilde{P}_{gg}^{(0)}(N) = 2C_A\left( 
\left[
\hat{N}_{+2} + \hat{N}_{-2}
 \right.
\right. 
\nn \\
&& \hspace{12.ex}
\left. \left. 
-2(\hat{N}_{+} + \hat{N}_- )
+3 
\right]S_2 
-\frac{\pi^2}{6}
\right)
\,.
\eea 

If we take the evolution of the $f_j(N+1,\mu_0) = {\cal D}_{jk}^{-1}(N+1) f_{k}(N+1,\mu)$ into account, we can derive the evolution for the matching coefficient $I_{ij}$ at NLL, which is 
\bea\label{eq:INLLevo} 
 I_{ij}(u\mu) &=&   
  {\cal D}^N_{ik}(\mu,\mu_0)
  \, I_{kl}(u\mu_0) {{\cal D}_{lj}^N}(\mu_0,\mu) \nn \\ 
 &&  
 \hspace{-10.ex} 
+  \frac{\alpha_s(\mu_0)}{2\pi} {\cal N}_{ik}  
\,
 [2P_{kl}^{(0)}(N)-2P_{kl}^{(0)}(N+1)]{\cal D}_{lj}^{N+1}(\mu_0,\mu) \,. \nn \\  
\eea 
We note that the analytic form for NLL we derived can be straightforwardly generalized to higher logarithmic accuracy. 

In practice, to implement the resummation, we  use the fact that $\sigma_{q_i}$ is identical to $\sigma_{{\bar q}_i}$ to re-cast the cross section as 
\bea
 \Sigma_N &=& \frac{4\pi\alpha^2}{Q^4} 
 \sum_{i=1}^{N_F} e_{q_i}^2 \left(
\hat{\sigma}_q \odot \frac{1}{N_F}(F^{NS}_{q_i}+F_q^S) 
+  2
\hat{\sigma}_g \odot F_g^S
\right) 
\nn \\ 
&& \hspace{-8.ex} =
\frac{1}{N_F} \frac{4\pi\alpha^2}{Q^4}
\sum_{i=1}^{N_F}
e_{q_i}^2
\hat{\sigma}_q
\odot F^{NS}_{q_i} 
+ \hat{\sigma}^S_q 
\odot F_q^S
+ \hat{\sigma}_g
\odot F_g^S \,.   
\eea 
Here we introduced $\hat{\sigma}_q^S = \frac{4\pi\alpha^2}{Q^4}\frac{1}{N_F} \sum_{i=1}^{N_F}e_{q_i}^2\hat{\sigma}_q $ and $\hat{\sigma}^S_g = \frac{4\pi\alpha^2}{Q^4} \frac{1}{N_F}\sum_{i=1}^{N_F}e_{q_i}^2(2N_F \hat{\sigma}_g)$. Inserting the resummed formula Eq.~(\ref{eq:fNLLevo}) for $F^{NS}$ and $F^S$, we realize the NLL resummation of $\Sigma_N(Q^2,\theta)$.  

%

\subsection{$d\Sigma_N/d\theta^2$-distribution at ${\cal O}(\alpha^2_s)$}\label{sec:lns}

The NLL resummation for $\Sigma_N$ allows us to predict the complete $d\Sigma_N/d\theta^2$ spectrum up to $\alpha_s^2$ order by expanding the resummation results in terms of the coupling $\alpha_s$. Here we list the results.

The distribution can be written as
\bea
\frac{1}{\sigma_0}\frac{d\Sigma_N}{d\ln\theta^2}
= \left(\frac{\alpha_s}{2\pi} \Sigma_{N,j}^{(1)} 
+\frac{\alpha_s^2}{4\pi^2}
\sum_{\vec{i}}\Sigma_{N,j}^{\vec{i}} \right) f_j(N+1) 
\eea 
where $\sigma_0 = \frac{4\pi\alpha^2e_q^2}{Q^4} $. Here $\vec{i} = (i_1,i_2,i_3,i_4)$ satisfies $i_1+i_2+i_3+i_4 = 2$ and $i_k\ge 0$. 

At ${\cal O}(\alpha_s)$, the distribution is given by 
\bea 
 \Sigma^{(1)}_{N,j}
=  
P_{qj}^{(0)}(N)-P_{qj}^{(0)}(N+1) \,.
\eea 

At ${\cal O}(\alpha_s^2)$, we have contributions coming from the $\alpha_s$ running, which are 
\bea 
\Sigma_{N,j}^{(0,0,1,1)} = -\ln\frac{\theta^2Q^2}{4\mu^2} \frac{\beta_0}{2}  
  (P_{qj}^{(0)}(N)-P_{qj}^{(0)}(N+1))     \,, 
\eea 
and 
\bea 
\Sigma_{N,j}^{(0,1,0,1)} = 
\frac{\beta_0}{2}   \,
  (d^{(1)}_{qj}(N) -d^{(1)}_{qj}(N+1)) \,.
\eea 
In addition, we have 
\bea 
\Sigma_{N,j}^{(1,0,1,0)}
 = \sigma^{(1)}_{i}(N)
 (P^{(0)}_{ij}(N)-P^{(0)}_{ij}(N+1))  \,, 
\eea 
is essentially the product of the ${\cal O}(\alpha_s)$ hard function in Section~\ref{sec:hard} and the ${\cal O}(\alpha_s)$ NEEC in Section~\ref{sec:collinear}. 

The $1$-loop DGLAP evolution contributes as
\bea 
\Sigma_{N,j}^{(0,0,2,0)}
 = 
 \Big(P^{(1)}_{qj}(N)-P^{(1)}_{qj}(N+1)\Big) \,. 
\eea 
Here, the moment of the NLO splitting function can be found in Ref.~\cite{Moch:2004pa,Vogt:2004mw} and is also provided in the Appendix.

The product of the LO DGLAP and the NLO matching coefficient contributes to both the 
double and single logs. The double logarithmic term reads  
\bea 
\Sigma_{N,a,j}^{(0,1,1,0)} 
&=&  -\ln\frac{\theta^2Q^2}{4\mu^2} \Big[ P_{qk}^{(0)}(N) 
(P_{kj}^{(0)}(N) -P_{kj}^{(0)}(N+1)   )  \nn \\
& -  &
(P_{qk}^{(0)}(N) -P_{qk}^{(0)}(N+1)   ) 
P_{kj}^{(0)}(N+1) \Big]  \,, 
\eea 
while the single log contribution is 
\bea 
\Sigma_{N,b,j}^{(0,1,1,0)} 
&=&   P_{qk}^{(0)}(N) (d_{kj}^{(1)}(N)-d_{kj}^{(1)}(N+1))   \nn \\
&& -  
(d_{qk}^{(1)}(N)- d_{qk}^{(1)}(N+1))
P_{kj}^{(0)}(N+1)  \,.
\eea 

There is one additional term that originated from the $\odot$ structure, which is 
single log term from $P^{(0)}$
\bea 
\Sigma_{N,c,j}^{(0,1,1,0)} 
=  2 \tilde{P}_{qk}^{(0)}(N) (P_{kj}^{(0)}(N) -P_{kj}^{(0)}(N+1) )  \,, 
\eea 
where $\tilde{P}^{(0)}_{ij}(N)$ is defined in Eq.~(\ref{eq:Ptilde}) and Eq.~(\ref{eq:Ptildeij}).

\section{Numerical Results}\label{sec:numerics}
In this section, we examine the numerical consequence of the NLL resummation. We use the kinematics that $E_P = 275\, {\rm GeV}$ for the incoming proton and $E_l = 18\, {\rm GeV}$ for the electron. We work in the Breit-frame and choose $N=3$, $Q^2=100{\rm GeV}^2$ and fix $\mu = Q$ and $\mu_0 = \frac{Q\theta}{2}$ for implementing resummation. 

First, we validate the factorization formalism by comparing the singular $\ln \theta$ contributions predicted by the factorization theorem with the complete $\alpha_s$ and $\alpha_s^2$ calculations of the distribution $d\Sigma_N/dy$, where $y = \ln(\tan\frac{\theta}{2})$. As $\theta$~($y$) becomes small, the $\ln \theta$ terms will dominate the $d\Sigma_N/dy$ distribution, and the singular contribution should coincide with the full calculation.
 \begin{figure}[htbp]
  \begin{center}
   \includegraphics[scale=0.168]{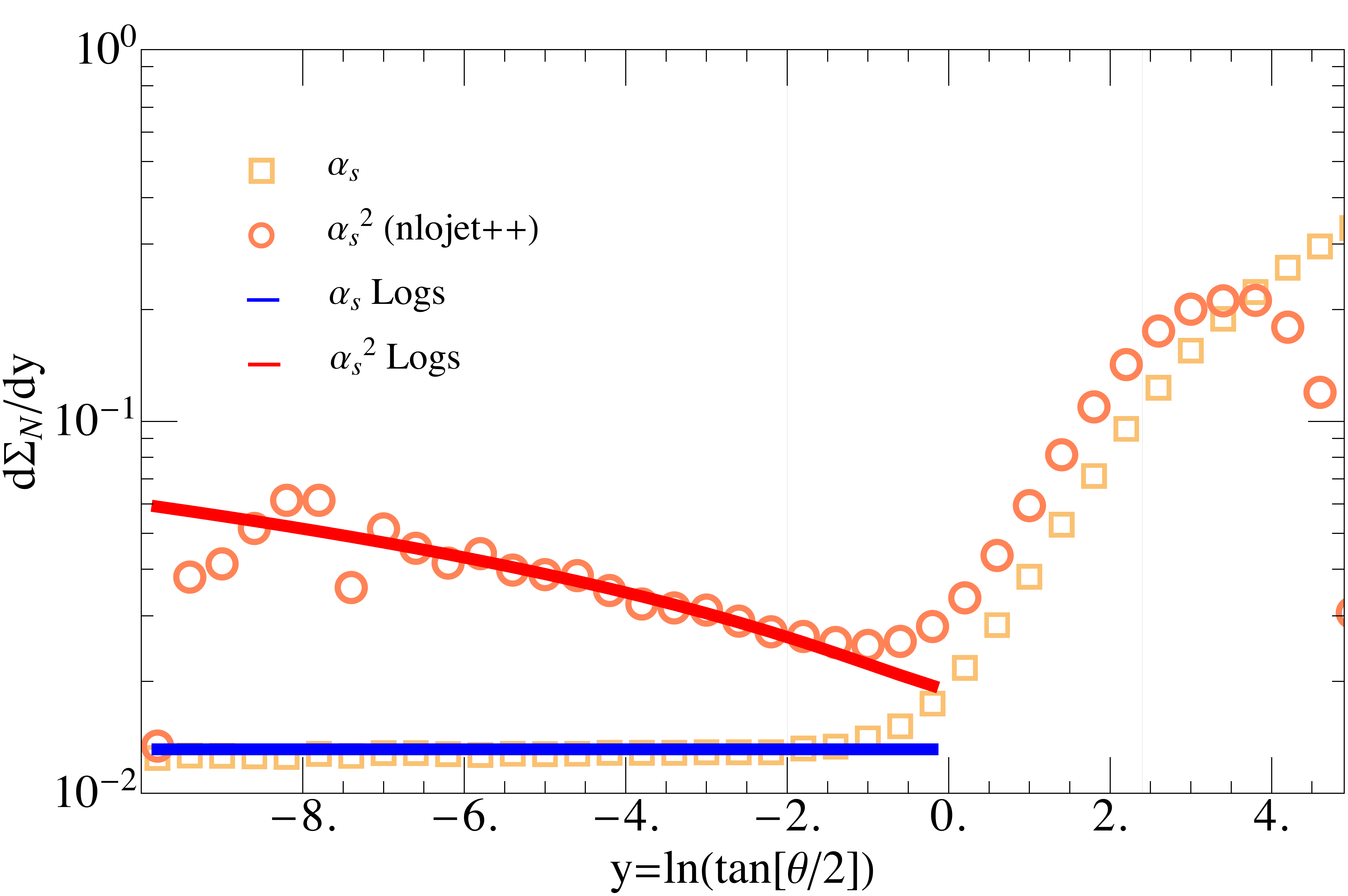} 
\caption{A comparison between the $\ln\theta$ singular contributions with the full fixed order calculations. Very good agreements are found for small values of $\theta$ ($y$).}
  \label{fg:singular}
 \end{center}
\end{figure}

We perform this comparison in fig.~\ref{fg:singular}.
The full fixed order calculations (in dots) are obtained numerically using  {\tt nlojet++}~\cite{Nagy:2005gn} and the log terms have been calculated in Section~\ref{sec:lns}. From fig.~\ref{fg:singular}, we observed very good agreements in the small $y$ region between the complete calculation and the singular terms predicted by factorization and resummation, in both the magnitude and shape. The comparison serves as a non-trivial validation of the factorization theorem derived in this work.

\begin{figure}[htbp]
  \begin{center}
   \includegraphics[scale=0.168]{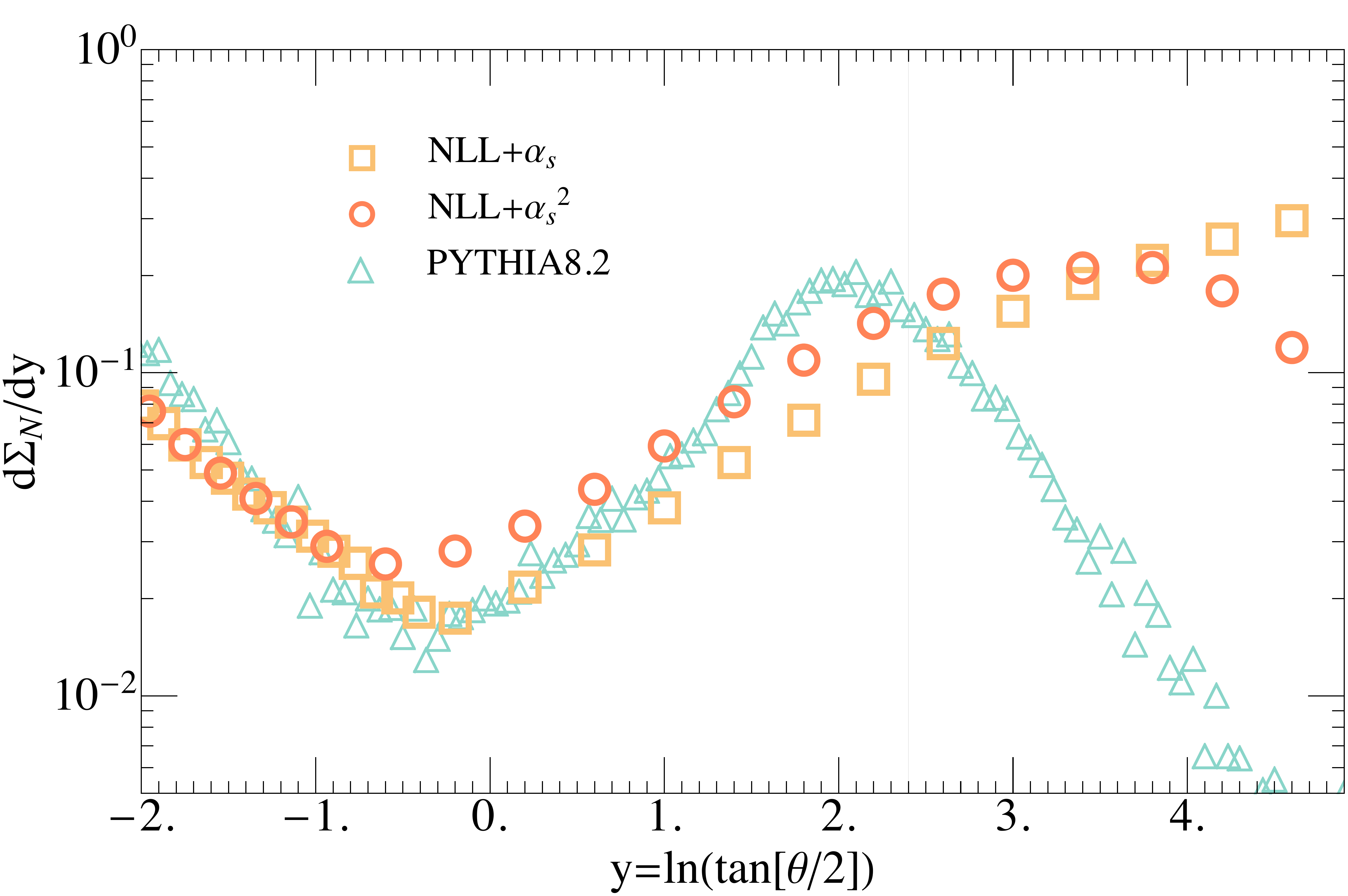} 
\caption{Comparison of the ${\rm NLL}+\alpha_s$, ${\rm NLL}+\alpha_s^2$ and the {\tt Pythia} simulation at partonic level. Reasonable agreement is found in the small $\theta$ ($y$) region (near-side) between the analytic NLL resummation and the {\tt Pythia} simulation. We stop the resummation at $y=-2$, after which one probes the non-perturbative region.  Additional TMD resummation is required for $\theta \to \pi$ (large $y$, away-side).}
  \label{fg:nll}
 \end{center}
\end{figure}

Now we present the numerical results for the resummation in fig.~\ref{fg:nll}. The analytic formula Eq.~(\ref{eq:fNLLevo}) is checked against the numerical resolution of Eq.~(\ref{eq:evo2}) using Euler's Method to find perfect agreement. 
We further matched the  resummation to the fixed order calculation by removing the singular terms that have been resummed,  from the fixed order cross section in the small $y$ region, and replacing them with the NLL results. In fig.~\ref{fg:nll}, We show the NLL$+\alpha_s$ and NLL$+\alpha_s^2$ in the orange square and red circular dots, respectively. Compared with the fixed order results in fig.~\ref{fg:singular}, we see that the resummation effects are significant in the small angle region, which enhances the distribution by several times with respect to the $\alpha_s^2$ calculation for $y$ around $-2$. It is also interesting to point out that as obvious in fig.~\ref{fg:nll}, the distribution in the small angle is not suppressed due to the absence of the Sudakov factor. This feature of the NEEC is very different from the TMD PDFs in which the small transverse momentum region is exponentially suppressed by the Sudakov factor. 

When $y<-2$, for the kinematics we chose, $\frac{Q\theta}{2}$ is comparable with $\Lambda_{\rm QCD}$ and we start to probe the non-perturbative region. The perturbative calculation is no longer valid in this regime and future experimental measurements at HERA or EIC are required to understand the non-perturbative behavior of the NEEC $f_{\rm EEC}(\theta)$ in this range, which in turn can teach us about the nucleon intrinsic transverse dynamics, as suggested by Ref.~\cite{Liu:2022wop}. 

We further compare the NLL+fixed order distributions with the {\tt Pythia8.2} simulation~\cite{Sjostrand:2014zea} which implements the leading logarithmic (LL) resummation. For this comparison, we have turned off hadronization in {\tt Pythia}. In the small $\theta$ ($y$) region (near-side region), the analytic NLL resummation agrees reasonably well with the partonic {\tt Pythia} simulation. For $0. <y<1.0$ ($\frac{\pi}{2}<\theta \lesssim 0.8\pi$), the ${\rm NLL}+\alpha_s$ agrees better with {\tt Pythia} and both are lower than the ${\rm NLL}+\alpha_s^2$ prediction, due to the missing higher order corrections in {\tt Pythia} and the ${\rm NLL}+\alpha_s$. For larger values of $y$ where $\theta$ is approaching $\pi$ (away-side region), the fixed order calculations differ substantially from the {\tt Pythia} simulation. In this region, the detected particles are almost back-to-back against the incoming proton. Now, the distribution becomes highly sensitive to the soft radiations (as well as the initial state collinear radiations), and we are essentially probing the small transverse momentum and therefore the TMD PDF. Therefore, in this region, additional TMD resummation is required for reliable predictions which we leave for future studies.

\section{Summary}\label{sec:summary}
In this work, we demonstrate the factorization theorem for the nucleon energy-energy correlator (NEEC) measurement in lepton-ion collisions proposed in~\cite{Liu:2022wop}. Our main results are presented in Eq.~(\ref{eq:fact-x}), where the energy-weighted cross-section $\Sigma_N(Q^2,\theta)$ is factorized into the partonic DIS cross section and the NEEC $f_{\rm EEC}(z,\theta)$. The operator definition of the NEEC is given by Eq.~(\ref{eq:fqx}) and Eq.~(\ref{eq:fgx}). 
The factorized form in the Mellin space can be found in Eq.~(\ref{eq:fact-N}). 
The factorization theorem has a similar structure to the DIS cross-section, except that the collinear PDF is replaced by the NEEC. Moreover, the factorization theorem can be easily generalized to the hadron-hadron collisions at the Large Hadron Collider (LHC) by appropriately substituting the PDF with the NEEC when similar measurements are performed. For instance, if the proton NEEC is measured in the prompt photon production in $pA$ collisions $pA\to\gamma + X$, then the factorization is the same as the inclusive photon production with the replacement of the proton PDF $f_{i/P}(z)$ with its corresponding NEEC $f_{i,{\rm EEC}}(z,\theta)$. 

When $Q\theta \gg \Lambda_{\rm QCD}$, we showed in Eq.~(\ref{eq:Ix}) that the $f_{\rm EEC}$ can be further matched onto the collinear PDF, with perturbatively calculable matching coefficients determined by the QCD splitting functions. In this region, the factorization formalism Eq.~(\ref{eq:Ix}) suggests that the $df_{\rm EEC}/d\theta$ describes the $\theta$ correlation between the out-going parton energy density and the initial incoming parton density. 
The factorization theorem is validated by the excellent agreements between the ${\cal O}(\alpha_s^2)$ prediction of the factorization and the complete NLO  calculation of $d\Sigma_N(Q^2,\theta)/d\theta^2$. The next-to-leading logarithmic (NLL) resummation has also been carried out for the NEEC. In the momentum space, the NEEC evolves in a similar way as the collinear PDFs. In this work, we focused more on the Mellin space evolution of the NEEC. We obtained a fully analytic solution to the evolution equation in Eq.~(\ref{eq:fNLLevo}) and supplied all the necessary ingredients for the NLL resummation. The analytic formula can be easily extended to higher logarithmic accuracy. The numerical evaluation of the NLL resummation is found to agree with the {\tt Pythia} simulation reasonably well. Furthermore, the NLL calculation also supports the recent idea of using the NEEC to look for the gluon saturation in lepton-ion collisions~\cite{Liu:2023aqb}, where the $\theta$ distribution predicted by the collinear factorization is not suppressed in the small $\theta$ region contrary to the expectation of the color glass condensate (CGC) effective framework. The NNLO calculation of the $f_{\rm EEC}$ in the perturbative region should be feasible with current computational techniques, which would enable us to perform NNLL resummation for the $f_{\rm EEC}$. We have not studied non-perturbative effects in this work and we plan to do so in future work. We hope our current work serves as a first step towards the precision measurement of $f_{\rm EEC}$ and stimulates further theoretical and experimental investigations.

\begin{acknowledgments}
We thank Jian-Hui Zhang for the useful discussions. This work is supported by the Natural Science Foundation of China under contract No.~12175016 (H.~C. and X.~L.), No.~11975200 (H.~X.~Z.) and  No.~12147103 (H.~X.~Z.).
\end{acknowledgments}

\begin{widetext} 
\appendix 
\section{solving the RG-evolution}\label{sec:derive}
In this section, we solve Eq.~(\ref{eq:evo2}), which can be written as
\bea\label{eq:inteform} 
f_{{\rm EEC}}(N,\ln\frac{Q\theta}{u \mu})  
=  
f_{{\rm EEC}}(N,\ln\frac{Q\theta}{u \mu_0})+ 
\int^\mu_{\mu_0} d\ln {\mu'}^2
  \int d\xi \xi^{N-1} 
P\left(\xi\right)
 f_{{\rm EEC}}(N,\ln\frac{Q\theta}{\xi \, u \mu'}) 
 \,.
\eea 
For simplicity, we have suppressed the subscripts. The product of the $P$'s should be treated as the matrix product.

We write the ansatz solution to Eq.~(\ref{eq:inteform}) as 
\bea 
f_{{\rm EEC}}(N,\ln\frac{Q\theta}{u \mu}) 
= D(\mu,\mu_0)f_{{\rm EEC}}(N,\ln\frac{Q\theta}{u \mu_0}) + R(\mu,\mu_0)\,,
\eea 
where $D$ and $R$ are to be determined and satisfy $D(\mu_0,\mu_0) = 1$ and $R(\mu_0,\mu_0)=0$. 

We plug the ansatz back into Eq.~(\ref{eq:inteform}), to find  
\bea\label{eq:start}
D(\mu,\mu_0)f_{{\rm EEC}}(N,\ln\frac{Q\theta}{u \mu_0})
+ R(\mu,\mu_0)
&=& f_{{\rm EEC}}(N,\ln\frac{Q\theta}{u \mu_0}) 
+\int_{\mu_0}^\mu d\ln{\mu'}^2 P(N,\mu') R(\mu',\mu_0)
\nn \\
&+ & 
\int^\mu_{\mu_0} d\ln {\mu'}^2
  \int d\xi \xi^{N-1} 
P\left(\xi\right) \, D(\mu',\mu_0)f_{{\rm EEC}}(N,\ln\frac{Q\theta}{\xi u \mu_0})\,. 
\eea 
To realize the NLL resummation, we use the NLO NEEC as the initial input at $\mu_0$, and manipulate Eq.~(\ref{eq:start}) as
\bea
D(\mu,\mu_0)f_{{\rm EEC}}(N,\ln\frac{Q\theta}{u \mu_0})
+R(\mu,\mu_0) 
&=& f_{{\rm EEC}}(N,\ln\frac{Q\theta}{u \mu_0}) \nn\\
&&\hspace{-15.ex} + \int^\mu_{\mu_0} d\ln {\mu'}^2
P\left(N,\mu'\right) \, \left[ D(\mu',\mu_0)f_{{\rm EEC}}(N,\ln\frac{Q\theta}{ u \mu_0}) 
+ R(\mu',\mu_0)
\right] 
\nn \\
&&\hspace{-15.ex} - \frac{\alpha_s(\mu_0)}{2\pi}
\int^\mu_{\mu_0} d\ln {\mu'}^2 
\tilde{P}\left(N,\mu'\right)  
D(\mu',\mu_0)
[2P(N)-2P(N+1)]f(N+1,\mu_0)
\,,
\eea 
where we have used the property that at NLO, the initial condition satisfies
\bea\label{eq:lnu} 
f_{\rm EEC}(N,\ln\frac{Q\theta}{\xi u \mu_0}) 
= f_{\rm EEC}(N,\ln\frac{Q\theta}{ u \mu_0}) 
-\frac{\alpha_s(\mu_0)}{2\pi}
\ln \xi 
[2P(N,\mu_0)-2P(N+1,\mu_0)]f(N+1,\mu_0)\,. 
\eea
and applied the definition 
\bea 
\tilde{P}(N) = \int_0^1 d\xi \xi^{N-1} P(\xi) \ln\xi \,. 
\eea 

Now we repeat the above procedure, to replace the $D(\mu',\mu_0)f_{\rm EEC}(N,\ln\frac{Q\theta}{\xi u \mu_0})+R(\mu',\mu_0)$ using Eq.~(\ref{eq:start}) to find 
\bea
D(\mu,\mu_0)f_{{\rm EEC}}(N,\ln\frac{Q\theta}{u \mu_0})
+ R(\mu,\mu_0)
&=& f_{{\rm EEC}}(N,\ln\frac{Q\theta}{u \mu_0})  
+ \int^\mu_{\mu_0} d\ln {\mu'}^2
P\left(N,\mu'\right) \, f_{{\rm EEC}}(N,\ln\frac{Q\theta}{ u \mu_0})
\nn \\
& &  \hspace{-20.ex}
+ \int_{\mu_0}^\mu d\ln{\mu'}^2 P(N,\mu') \int_{\mu_0}^{\mu'} d\ln{\mu''}^2 P(N,\mu'')R(\mu'',\mu_0)
\nn \\
&&\hspace{-20.ex}
+ \int^\mu_{\mu_0} d\ln {\mu'}^2
P\left(N,\mu'\right) \,
\int^{\mu'}_{\mu_0} d\ln {\mu''}^2\, 
P\left(N,\mu''\right) 
D(\mu'',\mu_0)f_{{\rm EEC}}(N,\ln\frac{Q\theta}{  u \mu_0})
\nn \\
&&\hspace{-20.ex}
- 
\frac{\alpha_s(\mu_0)}{2\pi}
\int^\mu_{\mu_0} d\ln {\mu'}^2
P\left(N,\mu'\right) \,
\int^{\mu'}_{\mu_0} d\ln {\mu''}^2
\tilde{P}\left(N,\mu''\right) 
D(\mu'',\mu_0)[2P(N)-2P(N+1)]f(N+1,\mu_0)
\nn \\
&&\hspace{-20.ex}
-\frac{\alpha_s(\mu_0)}{2\pi}
\int^\mu_{\mu_0} d\ln {\mu'}^2 
\tilde{P}\left(N,\mu'\right)  
D(\mu',\mu_0)
[2P(N)-2P(N+1)]f(N+1,\mu_0) \,, 
\eea 
which can be organized as 
\bea
&&D(\mu,\mu_0)f_{{\rm EEC}}(N,\ln\frac{Q\theta}{u \mu_0}) 
+ R(\mu,\mu_0)
\nn \\ 
&=& \int^\mu_{\mu_0} d\ln {\mu'}^2
P\left(N,\mu'\right) \, 
\int^{\mu'}_{\mu_0} d\ln {\mu''}^2\, 
P\left(N,\mu''\right) 
\left[ D(\mu'',\mu_0)
f_{{\rm EEC}}(N,\ln\frac{Q\theta}{u \mu_0})
+R(\mu'',\mu_0)
\right] \nn\\
&+& 
\left[1+ \int^\mu_{\mu_0} d\ln {\mu'}^2
P\left(N,\mu'\right)
\right]
f_{{\rm EEC}}(N,\ln\frac{Q\theta}{u \mu_0}) \nn\\
&-& 
\frac{\alpha_s(\mu_0)}{2\pi}
\int^{\mu}_{\mu_0} d\ln {\mu'}^2
\left[1+ \int^\mu_{\mu'} d\ln {\mu''}^2
P\left(N,\mu''\right) \right]\,
\tilde{P}\left(N,\mu'\right) 
D(\mu',\mu_0)[2P(N)-2P(N+1)]f(N+1,\mu_0)
\eea 
where in the last line, we have switched the order of the integrations, using 
\bea 
\int_{\mu_0}^\mu d \mu'  A(\mu') \int_{\mu_0}^{\mu'} d \mu'' B(\mu'') 
= \int_{\mu_0}^{\mu} d \mu'' B(\mu'')  \int_{\mu''}^\mu d \mu'  A(\mu')  \,.
\eea 

Iterate the procedure, we will arrive at  
\bea 
&& D(\mu,\mu_0) f_{\rm EEC}(N,\ln \frac{Q\theta}{u \mu_0} ) 
+R(\mu,\mu_0) \nn \\ 
&-& \lim_{n\to \infty} 
\int_{\mu_{n-1}}^\mu d\ln\mu_n^2 P(N,\mu_n) \dots \int^{\mu}_{\mu1} d\ln\mu_2^2 P(N,\mu_2)
\left[D(\mu_1,\mu_0) f_{\rm EEC}(N,\ln \frac{Q\theta}{u \mu_0} ) 
+R(\mu_1,\mu_0) \right]
\nn \\
&=& {\cal D}(\mu,\mu_0)f_{\rm EEC}(N,\ln\frac{Q\theta}{u\mu_0})
- \frac{\alpha_s(\mu_0)}{2\pi} 
\int_{\mu_0}^\mu d\ln\mu'^2
{\cal D}(\mu,\mu')\tilde{P}(N,\mu')D(\mu',\mu_0)[2P(N,\mu_0)-2P(N+1,\mu_0)]f(N+1,\mu_0) \,, \nn \\ 
\eea 
where ${\cal D} = \exp\left[\int_{\mu_0}^\mu d\ln\mu'^2 P(N,\mu') \right]$ is defined in Eq.~(\ref{eq:D}). 
We note that 
\bea 
&&
\lim_{n\to \infty} \left(\frac{ {\rm min}_{\mu} P(N,\mu)}{n}\right)^{n-1} f_{\rm EEC}(N,\ln\frac{Q\theta}{u\mu_0}) \to 0  
\nn \\ 
&& 
<
\lim_{n\to \infty} 
\int_{\mu_{n-1}}^\mu d\ln\mu_n^2 P(N,\mu_n) \dots \int^{\mu}_{\mu1} d\ln\mu_2^2 P(N,\mu_2)
\left[D(\mu_1,\mu_0) f_{\rm EEC}(N,\ln \frac{Q\theta}{u \mu_0} ) 
+R(\mu_1,\mu_0) \right] \nn \\
&&
< 
\lim_{n\to \infty} \left(\frac{ {\rm max}_{\mu} P(N,\mu)}{n}\right)^{n-1} f_{\rm EEC}(N,\ln\frac{Q\theta}{u\mu_0}) \to 0 \,. 
\eea 
Here we have assumed that the moment of the PDF is bounded and thus the limit vanishes as $n\to \infty$. 

Therefore, we conclude that 
\bea 
&&f_{\rm EEC}(N,\ln\frac{Q\theta}{u\mu})= D(\mu,\mu_0) f_{\rm EEC}(N,\ln \frac{Q\theta}{u \mu_0} ) 
+R(\mu,\mu_0) 
\nn \\ 
&=& {\cal D}(\mu,\mu_0)f_{\rm EEC}(N,\ln\frac{Q\theta}{u\mu_0})
- \frac{\alpha_s(\mu_0)}{2\pi} 
\int_{\mu_0}^\mu d\ln\mu'^2
{\cal D}(\mu,\mu')\tilde{P}(N,\mu')D(\mu',\mu_0)[2P(N,\mu_0)-2P(N+1,\mu_0)]f(N+1,\mu_0) \,. \nn \\ 
\eea

Since $f_{\rm EEC}(N,\ln\frac{Q\theta}{u\mu_0})$ and $R$ are independent and the solution should hold for arbitrary constant in $f_{\rm EEC}(N,\ln\frac{Q\theta}{u\mu_0})$, then we could identify 
\bea
&&D = {\cal D} = \exp\left[\int_{\mu_0}^\mu d\ln\mu'^2 P(N,\mu') \right]
\,, \nn \\
&& R = - \frac{\alpha_s(\mu_0)}{2\pi}
\int_{\mu_0}^\mu d\ln\mu'^2
 {\cal D}(\mu,\mu')\tilde{P}(N,\mu'){\cal D}(\mu',\mu_0)[2P(N,\mu_0)-2P(N+1,\mu_0)]f(N+1,\mu_0) \,.  
\eea

The derivation is applicable to higher logarithmic accuracy by suitably adjusting the relation in the initial condition of Eq.(~\ref{eq:lnu}) at higher $\alpha_s$ orders. 


\section{constant and function}
In this Appendix, we list the QCD color constants and splitting functions that are present in the main text. 

In QCD, the running of the strong coupling constant $\alpha_s$ follows 
\bea 
\frac{d\alpha_s}{d\ln\mu} 
= \beta[\alpha_s] \,, 
\eea 
where the $\beta$-function can be expanded in terms of $\alpha_s$ as
\bea 
\beta[\alpha_s] =- 2\alpha_s \sum \beta_n \left( \frac{\alpha_s} {4\pi} \right)^{n+1} \,,  
\eea 
with 
\bea 
\beta_0 = \frac{11}{3}C_A - \frac{4}{3}  N_F T_R \,, \qquad 
\beta_1 = \frac{34}{3}C_A^2
- \frac{20}{3}C_A T_R N_F - 4 C_F T_R N_F \,. 
\eea 
Here $C_A = N_C = 3$, $T_R = \frac{1}{2}$ and $C_F = \frac{N_C^2-1}{2N_C}$. $N_F$ is the number of quarks.

It is useful to note that~\cite{Li:2014ria} 
\bea 
\frac{\alpha_s(\mu) }{2\pi} - \frac{\alpha_s(\mu_0)}{2\pi}   = 
\frac{1}{2 }\frac{\alpha_s^2}{4\pi^2} \beta_0 \ln \frac{\mu_0^2}{\mu^2} + \dots
\,, \qquad 
\ln \left( \frac{\alpha_s(\mu) }{\alpha_s(\mu_0)} \right)   = 
 \frac{\beta_0}{2 }
 \frac{\alpha_s(\mu)}{2\pi}
  \ln \frac{\mu_0^2}{\mu^2} 
  + \frac{\alpha_s^2}{32\pi^2}
  \left(2\beta_1 \ln\frac{\mu_0^2}{\mu^2}
  -\beta_0^2 \ln^2 \frac{\mu_0^2}{\mu^2} \right)
  + \dots 
\eea 
and 
\bea\label{eq:ana1} 
\int_{\mu_0}^\mu \left(P_0\frac{\alpha_s}{2\pi}
+ P_1 \frac{\alpha_s^2}{4\pi^2}
\right)d\ln \mu^2 = 
- \frac{2}{\beta_0}
\left( P_0\ln\frac{\alpha_s(\mu)}{\alpha_s(\mu_0)}
+ (2 P_1-r_1P_0) \frac{\alpha_s(\mu)-\alpha_s(\mu_0)}{4\pi}
 + \dots  
\right)  
\eea 
where $r_i = \frac{\beta_i}{\beta_0}$. Here the first term on the right-hand starts from the LL ($\sim {\cal O}(\alpha_s L)$) and the second term contributes to the NLL ($\sim {\cal O}(\alpha_s L^2)$). 

Also, we have 
\bea\label{eq:ana2}  
\int_{\mu_0}^\mu d\ln {\mu'}^2 
\left(\frac{\alpha_s(\mu)}{\alpha_s(\mu')} \right)^{-\frac{2}{\beta_0}P_{ii}^{(0)}} 
\frac{\alpha_s(\mu')}{2\pi} \tilde{P}_{ij}^{(0)}
\left(\frac{\alpha_s(\mu')}{\alpha_s(\mu_0)} \right)^{-\frac{2}{\beta_0}P_{jj}^{(0)}} 
= \frac{\tilde{P}_{ij}^{(0)}}{P^{(0)}_{ii}-P_{jj}^{(0)}} 
\left[ 
\left(\frac{\alpha_s(\mu)}{\alpha_s(\mu_0)} \right)^{-\frac{2}{\beta_0}P^{(0)}_{ii}}
-\left(\frac{\alpha_s(\mu)}{\alpha_s(\mu_0)} \right)^{-\frac{2}{\beta_0}P^{(0)}_{jj}}
\right]
\eea 

The collinear splitting function $P_{ij}(z)$ that governs the PDF DGLAP evolution 
\bea 
\frac{d}{d\ln\mu^2} f_i(z,\mu) = P_{ij} \otimes f_j(z,\mu) \,, 
\eea 
can be written as the power series in $\alpha_s$, which reads
\bea 
P_{ij}(z) =  \sum_{L=0}
\left( \frac{\alpha_s}{2\pi} \right)^{L+1} P^{(L)}_{ij}(z) \,. 
\eea 

In practice, it is always useful to consider the singlet and the non-singlet splitting functions $P^{S}_{ij}$ for $i=q,g$, and $P_{NS}^+$. Here the singular splitting functions are defined as
\bea 
&& P_{qq}^S = P_{NS}^+ + P_{ps}\,, \qquad 
P_{gg}^S = P_{gg} \,, \nn \\
&& P_{qg}^S = 2N_F P_{qg}\,, \qquad 
P_{gq}^S = P_{gq} \,. 
\eea 

In the $z$-space, at the LO
\bea 
&& P_{NS}^{+,(0)}(z) =  P_{qq}^{(0)}(z) = C_F \left(\frac{1+z^2}{1-z} \right)_+\,, \qquad 
P_{ps}^{(0)} = 0\,, \nn \\ 
&& 
P_{gq}^{S,(0)} = 
P_{gq}^{(0)}(z) = C_F \frac{1+(1-z)^2}{z} \,, \qquad 
P_{qg}^{S,(0)} = 2N_F P_{qg}^{(0)}(z) = 2N_F T_R (z^2+(1-z)^2) \,, 
\nn \\ 
&& 
P_{gg}^{S,(0)} = 
 P_{gg}^{(0)}(z) = 2C_A\left( 
\frac{z}{(1-z)_+} + \frac{1-z}{z}
+z(1-z) 
\right) + \frac{\beta_0}{2}\delta(1-z) \,. 
\eea 

In the Mellin space, we have 
\bea 
P_{NS}^{+,(0)} = C_F \left(\frac{3}{2} -(\hat{N}_++\hat{N}_-)S_1 \right)
\eea 
and 
\bea 
P_{NS}^{+,(1)} &=&  C_FC_A 
\left(- 2\hat{N}_+ S_3 + \frac{17}{24} 
+2 S_{-3} + \frac{28}{3} S_1
-(\hat{N}_- + \hat{N}_+) 
\left[ 
\frac{151}{18}S_1 + 2S_{1,-2} - \frac{11}{6}S_2
\right]
\right) \nn \\ 
&& + C_F N_F\left( 
-\frac{1}{12}-\frac{4}{3}S_1 
+(\hat{N}_-+\hat{N}_+) \left[\frac{11}{9}S_1-\frac{S_2}{3} \right]
\right) 
+C_F^2 \left( 
-4S_{-3}-2S_1-2S_2+\frac{3}{8}
\right.  \nn \\ 
&& \left.  - \hat{N}_-[S_2+2S_3] 
+ (\hat{N}_- + \hat{N}_+) 
\left[S_1+4S_{1,-2}+2S_{1,2}+2S_{2,1}+S_3 \right]
\right) 
\eea 

\bea 
P_{ps}^{(0)}(N) = 0\,,  
\eea 

\bea 
P_{ps}^{(1)}(N) &=& C_F N_F 
\left(
\frac{20}{9}(\hat{N}_- -\hat{N}_{-2})S_1
+(\hat{N}_+ - \hat{N}_{+2})\left[\frac{56}{9}S_1
+\frac{8}{3}S_2\right]
+ (\hat{N}_+ - 1) \left[8S_1 -4S_2\right] \right. \nn \\
&& \left. 
+ (\hat{N}_- - \hat{N}_+) \left[2S_1+S_2+2S_3 \right]
\right) 
\,,  
\eea

\bea 
P_{qg}^{S,(0)} = 2N_F P_{qg}^{(0)} 
=2 N_F T_R (-\hat{N}_- - 4\hat{N}_+ + 2 \hat{N}_{+2} +3 )S_1 \,.  
\eea 

\bea 
P_{qg}^{S,(1)} &=&  -C_A N_F \left( 
\frac{20}{9}({\hat N}_{-2}-{\hat N}_{-} )S_1
-({\hat N}_{-}-{\hat N}_{+})[2S_1+S_2+2S_3] \, 
-({\hat N}_{+}-{\hat N}_{+2}) \left[ 
\frac{218}{9}S_1 + 4S_{1,1}
+ \frac{44}{3}S_2 
\right] \right. \nn \\ 
&& \hspace{10.ex} 
\left. 
+(1-{\hat N}_{+})\left[27S_1 + 4S_{1,1}-7S_2-2S_3 \right]
-2({\hat N}_{-}+4{\hat N}_{+}-2{\hat N}_{+2}-3) \left[ S_{1,-2}+S_{1,1,1}\right]
\right)  \nn \\ 
&& - C_F N_F \left( 
2({\hat N}_{+} -{\hat N}_{+2})
\left[5S_1 + 2 S_{1,1}-2S_2+S_3 \right]
-(1-{\hat N}_{+})
\left[\frac{43}{2}S_1 + 4S_{1,1}-\frac{7}{2}S_2 \right] \right. \nn \\ 
&& \hspace{10.ex} 
\left. 
+ ({\hat N}_{-} - {\hat N}_{+})
\left[  7S_1 - \frac{3}{2}S_2 \right] 
+ 2({\hat N}_{-}+4{\hat N}_{+}-2{\hat N}_{+2}-3) 
\left[S_{1,1,1}-S_{1,2}-S_{2,1} +\frac{1}{2}S_3 
\right]
\right)  \,. 
\eea 

\bea 
P_{gq}^{S,(0)} = P_{gq}^{(0)} = C_F
(-2{\hat N}_{-2}+4{\hat N}_{-}+{\hat N}_{+}-3)S_1 \,.
\eea 

\bea 
 P_{gq}^{S,(1)} &=& -C_AC_F 
\left(  2(2{\hat N}_{-2}-4{\hat N}_{-}-{\hat N}_{+}+3)
\left[S_{1,1,1}-S_{1,-2}-S_{1,2}-S_{2,1} \right] 
+ (1-{\hat N}_{+})\left[2S_1-13S_{1,1}-7S_2-2S_3 \right]
\right.  \nn \\ 
&& \hspace{5.ex}
\left. 
+({\hat N}_{-2}-2{\hat N}_{-}+{\hat N}_{+}) \left[ S_1 - \frac{22}{3}S_{1,1}\right]
+4({\hat N}_{-}-{\hat N}_{+})\left[\frac{7}{9}S_1+3S_2+S_3 \right]
+({\hat N}_{+}-{\hat N}_{+2})\left[\frac{44}{9}S_1+\frac{8}{3}S_2 \right] \right) \nn \\ 
&&  
- C_F N_F  \left( 
({\hat N}_{-2}-2{\hat N}_{-}+{\hat N}_{+})
\left[\frac{4}{3}S_{1,1}-\frac{20}{9}S_1 \right]
-(1-{\hat N}_{+})\left[4S_1-2S_{1,1} \right]
\right) \nn \\ 
&& 
-C_F^2\left( 
(2{\hat N}_{-2}-4{\hat N}_{-}-{\hat N}_{+}+3)
\left[3S_{1,1}-2S_{1,1,1} \right]
-(1-{\hat N}_{+})\left[S_1-2S_{1,1}+\frac{3}{2}S_2-3S_3 \right] \right. \nn \\ 
&& \hspace{5.ex}
\left. - ({\hat N}_{-}-{\hat N}_{+})
\left[\frac{5}{2}S_1+2S_2+2S_3 \right]
\right) \,. 
\eea 

\bea 
P_{gg}^{S,(0)} = P_{gg}^{(0)} = 
2C_A(-{\hat N}_{-2}+2{\hat N}_{-}+2{\hat N}_{+}-{\hat N}_{+2}-3)S_1 + \frac{\beta_0}{2} \,, 
\eea 

\bea 
P_{gg}^{S,(1)} &=& - C_A N_F
\left(\frac{2}{3}-\frac{16}{3}S_1
-\frac{23}{9}({\hat N}_{-2}+{\hat N}_{+2})S_1
+\frac{14}{3}({\hat N}_{-}+{\hat N}_{+})S_1
+ \frac{2}{3}({\hat N}_{-}-{\hat N}_{+})S_2 
\right) \nn \\ 
&& - C_A^2 \left( 2 S_{-3} - \frac{8}{3} 
-\frac{14}{3}S_1 + 2S_3 
-4({\hat N}_{-2}-2{\hat N}_{-}-2{\hat N}_{+}+{\hat N}_{+2}+3)\left[S_{1,-2}+S_{1,2}+S_{2,1} \right] 
\right. \nn \\ 
&& \hspace{5.ex}\left. 
+ \frac{8}{3}({\hat N}_{+}-{\hat N}_{+2})S_2 
- 4({\hat N}_{-}-3{\hat N}_{+}+{\hat N}_{+2}+1)
\left[3S_2-S_3\right] +\frac{109}{18}({\hat N}_{-}+{\hat N}_{+})S_1 
+ \frac{61}{3}({\hat N}_{-}-{\hat N}_{+})S_2
\right)  \nn \\ 
&& - C_FN_F\left( 
\frac{1}{2} 
+ \frac{2}{3}
({\hat N}_{-2}-13{\hat N}_{-}-{\hat N}_{+}
-5{\hat N}_{+2}+18)S_1
+(3{\hat N}_{-}-5{\hat N}_{+}+2)S_2 
-2({\hat N}_{-}-{\hat N}_{+})S_3 
\right) \,. 
\eea

\section{partonic cross section in the $z$-space}
Here we list the DIS partonic cross section in the $z$-space, which can be written as 
\bea 
\hat{\sigma}(z,Q^2) 
=\frac{4\pi\alpha^2}{Q^4} \sum_{i=-N_F}^{N_F}\sum_{c=q,g}\sum_{\lambda = T,L} e^2_{q_i} f_\lambda \,  \hat{\sigma}_{c,\lambda}(z)
\eea 
where $\hat{\sigma}_{c,\lambda}(z)$ can be expanded in power series of the strong coupling constant $\alpha_s$,
\bea 
\hat{\sigma}_{c,\lambda}(z) = \sum_{n=0} \left(\frac{\alpha_s}{2\pi}\right)^n \hat{\sigma}^{(n)}_{c,\lambda} (z) \,.
\eea 
 At LO
\bea 
\hat{\sigma}^{(0)}_{q,T} = \delta(1-z) \,, \quad 
\hat{\sigma}^{(0)}_{q,L} = \hat{\sigma}^{(0)}_{g,T} =\hat{\sigma}^{(0)}_{g,L} = 0 \,.  
\eea 
The $\hat{\sigma}^{(1)}$'s have been known for a long time. The quark cross section reads 
\bea
\hat{\sigma}^{(1)}_{q,L}(z) 
=  C_F z\,,   
\eea 
for the longitudinal part, 
and 
\bea
&& \hat{\sigma}^{(1)}_{q,T}(z) =  C_F 
\Bigg\{ 
\left( 
\frac{1+z^2}{1-z}
\right) \left( \ln \frac{Q^2}{\mu^2} 
+ 
 \ln\frac{1-z}{z} \right) 
 -\frac{3}{2(1-z)} 
 +3 
 -\delta(1-z)\left(\frac{9}{2} + \frac{\pi^2}{3}\right)
\Bigg\} \,, 
\eea 
where $+$-distributions to regulate all 
divergences for $z \to 1$ are implied.

The gluon channel is given by 
\bea 
&&\hat{\sigma}_{L,g}^{(1)} =   T_R 
\left[2z(1-z) \right] \,, 
\eea 
and 
\bea 
&& \hat{\sigma}^{(1)}_{T,g} =  T_R 
\Bigg\{
(1-2z+2z^2)\ln\frac{Q^2}{\mu^2} + (1-2z+2z^2)\ln\frac{1-z}{z}
-1+4z(1-z) 
\Bigg\} \,.  
\eea

\section{useful formulae}
The following formulae are useful for deriving the Mellin transformation
\bea 
&& S_2(\infty) = \frac{\pi^2}{6} \,, \nn \\ 
&& \int_0^1 dz z^m \ln z = \frac{-1}{(m+1)^2} \,, \nn \\
&& \int_0^1 dz z^m \ln(1-z) = \frac{-1}{(m+1)}S_1(m+1) \,, \nn \\
&& \int_0^1 dz z^{N-1} \left( \frac{\ln(1-z)}{1-z} \right)_+
= \hat{N}_- S_{1,1} \,, \\ 
&& \int_0^1 dz z^{N-1} \frac{z}{1-z} \ln z 
= S_2 - \frac{\pi^2}{6} \nn \\ 
&& \int_0^1 dz z^{N-1} \frac{1-z}{z} \ln z 
= (\hat{N}_{-2}-2\hat{N}_-+1)S_2  \nn \\ 
&& \int_0^1 dz z^{N-1} z^m \ln z = [\hat{N}_{+m-1}-\hat{N}_{+m}] S_2 \,, \qquad m\in N  \nn  
\eea

\end{widetext}

\bibliographystyle{h-physrev}   
\bibliography{refs}

\end{document}